\documentclass[twocolumn,twocolappendix]{aastex63}

\usepackage[caption=false]{subfig}
\usepackage{xfrac}
\usepackage{slantsc}
\usepackage{chngcntr}
\usepackage[T1]{fontenc}
\newcommand{\HI}{H\textsc{i}}
\newcommand{\Msolar}{M$_{\odot}$}
\newcommand{\kms}{km\,s$^{-1}$}

\submitjournal{AJ}

\shorttitle{Faint tails in the Virgo cluster}
\shortauthors{R. Taylor et al.}

\begin{document}

\title{Faint and fading tails : the fate of stripped HI gas in Virgo cluster galaxies}

\correspondingauthor{Rhys Taylor}
\email{rhysyt@gmail.com}

\author[0000-0002-3782-1457]{Rhys Taylor}
\affiliation{Astronomical Institute of the Czech Academy of Sciences,\\
Bocni II 1401/1a, \\
141 00 Praha 4,
Czech Republic}

\author{Joachim K\"{o}ppen}
\affiliation{Astronomical Institute of the Czech Academy of Sciences,\\
Bocni II 1401/1a, \\
141 00 Praha 4,
Czech Republic}
\affiliation{Institut f\"{u}r Theoretische Physik und Astrophysik der Universit\"{a}t zu Kiel,\\
D-24098, Kiel, Germany.}

\author[0000-0002-1640-5657]{Pavel J\'{a}chym}
\affiliation{Astronomical Institute of the Czech Academy of Sciences,\\
Bocni II 1401/1a, \\
141 00 Praha 4,
Czech Republic}

\author[0000-0002-1261-6641]{Robert Minchin}
\affiliation{Stratospheric Observatory for Infrared Astronomy/USRA,\\
NASA Ames Research Center, MS 232-12,\\
Moffett Field, CA 94035, USA}

\author[0000-0001-6729-2851]{Jan Palou\v{s}}
\affiliation{Astronomical Institute of the Czech Academy of Sciences,\\
Bocni II 1401/1a, \\
141 00 Praha 4,
Czech Republic}

\author[0000-0003-1848-8967]{Richard W\"{u}nsch}
\affiliation{Astronomical Institute of the Czech Academy of Sciences,\\
Bocni II 1401/1a, \\
141 00 Praha 4,
Czech Republic}

%\nocollaboration{5}

% Example for quoting solar masses
% $\times$10$^{9}$\Msolar{}

\begin{abstract}
Although many galaxies in the Virgo cluster are known to have lost significant amounts of \HI{} gas, only about a dozen features are known where the \HI{} extends significantly outside its parent galaxy. Previous numerical simulations have predicted that \HI{} removed by ram pressure stripping should have column densities far in excess of the sensitivity limits of observational surveys. We construct a simple model to try and quantify how many streams we might expect to detect. This accounts for the expected random orientation of the streams in position and velocity space as well as the expected stream length and mass of stripped \HI{}. Using archival data from the Arecibo Galaxy Environment Survey, we search for any streams which might previously have been missed in earlier analyses. We report the confident detection of ten streams as well as sixteen other less sure detections. We show that these well-match our analytic predictions for which galaxies should be actively losing gas, however the mass of the streams is typically far below the amount of missing \HI{} in their parent galaxies, implying that a phase change and/or dispersal renders the gas undetectable. By estimating the orbital timescales we estimate that dissolution rates of 1-10\,\Msolar{}\,yr$^{-1}$ are able to explain both the presence of a few long, massive streams and the greater number of shorter, less massive features.
\end{abstract}

\section{Introduction}
\label{sec:intro}
It is well-established that many late-type galaxies in Virgo are strongly deficient in \HI{} : that is, they possess less \HI{} gas than similar field galaxies (e.g. \citealt{haynes86}, \citealt{solanes01}, \citealt{gv08}). In some cases this is equivalent to a loss of $>$\,6$\times$10$^{9}$\Msolar{} (\citealt{me12}). It is also well-known that a few galaxies in the cluster are associated with spectacular \HI{} streams up to 500 kpc in extent (\citealt{k08}), while others appear to have much shorter features (\citealt{chung}). There appears to be little or no correlation between which galaxies are deficient and which possess streams. Some of the longest streams are associated with galaxies which are even gas rich, whereas many strongly deficient galaxies apparently lack streams entirely.

The dominant mechanism for gas loss in clusters is thought to be ram pressure stripping (e.g. \citealt{guns}, \citealt{vol01}, \citealt{pavel1}, \citealt{roe08}, \citealt{pavel2}, \citealt{koppen}, for a detailed review see \citealt{bg06}). This can explain the complete removal of the gas content of a massive galaxy in a few orbits, whilst leaving the stellar component largely unaffected. In contrast, tidal encounters (e.g. \citealt{tooms}, \citealt{bekki}, \citealt{duc08}) have been shown to more likely result in only small amounts of gas being removed (e.g. \citealt{me17}).

Deficiency alone does not necessarily mean a galaxy is currently losing gas - it may have been stripped in the distant past, and even if close to the cluster centre (where ram pressure is expected to be strongest) in projection, its true 3D distance may be significantly greater. Detecting short streams is hampered by the low resolution of single-dish observations, with limited data available from interferometers with the necessary sensitivity. Physically, once gas is removed from its parent galaxy it may disperse into a larger volume, and/or it might experience a phase change (either by heating or cooling) and so rendered undetectable.

Yet collectively, the discrepancy between the numbers of deficient galaxies and those with streams seems too strong to ignore. The Arecibo-based ALFALFA (Arecibo Legacy Fast ALFA survey; \citealt{alfalfa}) and AGES (Arecibo Galaxy Environment Survey; \citealt{auld}) projects have both covered parts of the cluster to high sensitivity at 17 kpc resolution. Galaxies of low and high deficiency are found in close proximity to one another, strongly suggesting that at least some galaxies in the surveyed regions should be in the process of actively losing gas (\citealt{me12}, see also phase-space investigations, e.g. \citealt{jaff}, \citealt{rhee}). While the earlier VLA Imaging of Virgo in Atomic gas survey (VIVA, \citealt{chatlas}) discovered several short streams, both Arecibo surveys have reported few new \HI{} streams in Virgo - none at all in the case of AGES. 

In principle, a sufficiently rapid phase change could explain the dearth of streams (e.g \citealt{boss19}). Yet the existence of a few extremely long streams at least shows that this cannot be the complete explanation. If evaporation does account for the lack of most of the expected streams, why are there any long streams at all - especially near the centre of the cluster ? Why is there no correlation between the deficiency of a galaxy and the presence of a stream ? 

This problem has been remarked upon in \cite{oo05}, \cite{vol07}, and ourselves in \cite{me16}. \cite{oo05} describe a particularly interesting stream - it is approximately 100 kpc long, located near the centre of the Virgo cluster, and its length suggests a survival time $\sim$\,100 Myr. Its parent galaxy is strongly \HI{} deficient, though the mass of \HI{} in the plume can only account for about 10\% of the missing gas. Both molecular and ionised gas were later detected in the stream, but both are an order of magnitude less massive than the \HI{} (\citealt{verd}). In contrast, NGC 4569 is strongly \HI{} deficient but \cite{boss16} find that 17-42\% of its missing \HI{} can be explained by a phase change to (detected) H$\alpha$. It is therefore unclear if phase changes can explain the lack of \HI{} streams in the Virgo cluster.

In contrast, numerous ram-pressure stripping simulations have shown that stripped \HI{} can remain detectable at distances $>$\,100 kpc from its parent galaxy (e.g. \citealt{roe08}, \citealt{ton10}). Cosmological simulations by \cite{yun} found that 30\% or more of gas-rich galaxies possess long one-sided gas tails, though this does not account for the phase of the gas. Other simulations have shown that while harassment cannot explain strong deficiencies, it can still produce easily detectable, long, one-sided \HI{} streams (\citealt{duc08}, \citealt{me17}).

\cite{vol07} proposed that only the warm \HI{} is stripped by ram pressure, quickly rendering it undetectable due to evaporation and dispersal. But this would still allow a 200 Myr detectability window, and some galaxies are so deficient that it appears that the cold, inner \HI{} has also been stripped. In short, none of the scenarios proposed are very satisfying solutions to the `missing stream' problem - at least in Virgo.

In this paper we attempt to address these issues. Section \ref{sec:knownstreams} reviews the known optically dark \HI{} features in Virgo. In section \ref{sec:predictedstreams} we attempt to quantify how many streams we expect to detect. In section \ref{sec:searching} we use new analysis techniques to re-search AGES data cubes, uncovering a number of streams that were previously missed. We interpret the validity and physical nature of our detections in section \ref{sec:streamint}. Finally in section \ref{sec:conc} we comment on the these results and whether they alleviate the problems discussed above.

\section{Known optically dark \HI{} features in Virgo}
\label{sec:knownstreams}
In order to quantify the (possible) discreprancy between the actual and expected number of \HI{} streams in Virgo, we require a catalogue of known features. We compile a catalogue of optically dark \HI{} features in Virgo with $v_{hel}$\,$<$\,3,000\,\kms{} from a literature search (see also \citealt{me16} for more details), which is presented in table \ref{tab:virgogas}.

\begin{table*}
\centering
\caption{Properties of known optically dark \HI{} features in the Virgo cluster. The `name' column gives the parent galaxy (where available). Spatial coordinates are in J2000. All parameters, except coordinates, refer to the optically dark gas and not the parent galaxy (if one is present). We divide the table into three categories, separated by horizontal lines : the uppermost section contains discrete clouds, the middle section short streams, and the lower section long streams. Reference codes are as follows~: T12 = \cite{me12}; T13 = \cite{me13}; K09 = \cite{kent09}; K10 = \cite{kent10}; C07 = \cite{chung}; D07 = \cite{duc07}; M07 = \cite{m07}; G89 = \cite{ghcloud}; O05 = \cite{oo05}; K08 = \cite{k08}, S17 = \cite{sorgho}. The parent galaxy of VIRGOHI21 is believed to be NGC 4254 (VCC 307) while the Koopmann stream is associated with NGC 4534/DDO 137.}
\label{tab:virgogas}
\begin{tabular}{c c c c c c c c c c}
\hline
  \multicolumn{1}{c}{Name} &
  \multicolumn{1}{c}{RA} &
  \multicolumn{1}{c}{Dec} &
  \multicolumn{1}{c}{Velocity} &
  \multicolumn{1}{c}{Distance} &
  \multicolumn{1}{c}{M$_{\HI{}}$} &
  \multicolumn{1}{c}{Projected size} &
  \multicolumn{1}{c}{W50} &
  \multicolumn{1}{c}{W20} &
  \multicolumn{1}{c}{Reference} \\
  \multicolumn{1}{c}{} &
  \multicolumn{1}{c}{} &
  \multicolumn{1}{c}{} &
  \multicolumn{1}{c}{[\kms{}]} &
  \multicolumn{1}{c}{[Mpc]} &
  \multicolumn{1}{c}{[\Msolar{}]} &
  \multicolumn{1}{c}{[kpc]} &
  \multicolumn{1}{c}{[\kms{}]} &
  \multicolumn{1}{c}{[\kms{}]} &
  \multicolumn{1}{c}{} \\  
\hline
  AGESVC1 231 & 12:18:17.9 & 07:21:40 & 191 & 32 & 4.2E7 & $<$32 & 36 & 152 & T12\\
  AGESVC1 247 & 12:24:59.2 & 08:22:38 & 1087 & 23 & 2.3E7 & $<$23 & 22 & 33 & T12\\
  AGESVC1 257 & 12:36:55.1 & 07:25:48 & 1580 & 17 & 1.4E7 & $<$17 & 131 & 157 & T12\\
  AGESVC1 258 & 12:38:07.2 & 07:30:45 & 1786 & 17 & 1.4E7 & $<$17 & 32 & 120 & T12\\
  AGESVC1 262 & 12:32:27.2 & 07:51:52 & 1322 & 23 & 2.0E7 & $<$23 & 104 & 146 & T12\\
  AGESVC1 266 & 12:36:06.5 & 08:00:07 & 1691 & 17 & 3.2E7 & $<$17 & 77 & 173 & T12\\
  AGESVC1 274 & 12:30:25.6 & 08:38:05 & 1297 & 17 & 7.3E6 & $<$17 & 22 & 35 & T12\\
  AGESVC1 282 & 12:25:24.1 & 08:16:54 & 943 & 23 & 4.4E7 & $<$23 & 69 & 164 & T12\\
  AAK2 C1N & 12:08:47.6 & 11:55:57 & 1234 & 17 & 2.0E7 & 5 & 22 &  & K10\\
  AAK2 C1S & 12:08:47.4 & 11:54:48 & 1225 & 17 & 2.7E7 & 7 & 20 &  & K10\\
  AAK2 C2N & 12:13:42.5 & 12:54:50 & 2237 & 32 & 3.4E7 & 23 & 13 &  & K10\\
  AAK2 C2W & 12:13:33.1 & 12:52:44 & 2205 & 32 & 6.0E7 & 22 & 41 &  & K10\\
  AAK2 C2S & 12:13:41.9 & 12:51:16 & 2234 & 32 & 1.2E7 & 8 & 6 &  & K10\\
  VCC 1249 & 12:29:54.4 & 07:58:05 & 475 & 17 & 5.1E7 & $<$17 & 29 & 63 & T12\\
  KW cloud & 12:28:34.4 & 09:18:33 & 1270 & 17 & 7.0E7 & 37 & 73 &  & S17\\
\hline  
  VCC 865 & 12:25:58.8 & 15:40:17 & -128 & 17 & 7.0E7 & 13 & 110 &  & C07\\
  VCC 497 & 12:21:42.5 & 14:35:54 & 1149 & 17 & 9.0E7 & 16 & 50 &  & C07\\
  VCC 465 & 12:21:17.8 & 11:30:38 & 355 & 17 & 2.0E8 & 27 & 110 &  & C07\\
  VCC 1516 & 12:21:40.5 & 11:30:00 & 237 & 17 & 2.6E8 & 14 & 90 &  & C07\\
  VCC 630 & 12:23:17.2 & 11:22:05 & 1563 & 17 & 4.0E7 & 13 & 50 &  & C07\\
  VCC 2066 & 12:47:59.9 & 10:58:15 & 1141 & 17 & 9.0E8 & 50 & 78 &  & D07\\
  VCC 979 & 12:27:11.6 & 09:25:14 & 437 & 23 & 4.3E7 & 60 & 50 &  & S17\\
  VCC 1987 & 12:43:56.6 & 13:07:36 & 1046 & 17 & 4.1E8 & 32 & 90 &  & C07\\
\hline
  AA Virgo 7 & 12:30:25.8 & 09:28:01 & 488 & 17 & 5.3E8 & 173 & 127 &  & K09\\
  VIRGOHI21 & 12:17:52.9 & 14:47:19 & 2005 & 17 & 1.8E8 & 250 & 463 &  & M07\\
  HI1225+01 & 12:27:46.3 & 01:36:01 & 1292 & 17 & 2.9E9 & 182 & 60 &  & G89\\
  VCC 836 plume & 12:25:46.7 & 12:39:44 & 2524 & 17 & 3.8E8 & 124 & 550 &  & O05\\
  Koopmann & 12:34:19.3 & 06:28:04 & 2012 & 17 & 4.0E8 & 500 & 290 &  & K08\\
\hline
\end{tabular}
\end{table*}

We have arranged the features in table \ref{tab:virgogas} into three categories : isolated clouds; short streams attached to their parent galaxies; and much longer features. While most streams and clouds are typically less than 30 kpc in length, a handful are truly enormous, from 100-500 kpc in extent. Table \ref{tab:virgoggalas} compares how the streams relate to their parent galaxies. The \HI{} deficiency is quantified using the method of \cite{haynes84}, using the parameters of \citealt{solanes} (see their table 2). The intrinsic scatter in the relation is generally taken to be around 0.3.

\begin{table}
\centering
\caption{Galaxies with known \HI{} streams, comparing the parent galaxy \HI{} deficiency with the tail mass. We use the size and morphology parameters from GOLDMine to compute the expected \HI{} mass, using the method of \cite{solanes}. Tail masses are taken from the references in table \ref{tab:virgogas}. The missing mass is computed as the difference between the actual mass in the galaxy (ignoring the tail) and its expected mass. For VCC 1249 see section \ref{sec:VCC1249}. VCC 307 (NGC 4254) is the likely parent galaxy of the VIRGOHI21 feature.}
\label{tab:virgoggalas}
\begin{tabular}{c c c c c}
\hline
  \multicolumn{1}{c}{Name} &
  \multicolumn{1}{c}{M$_{\textrm{\HI{}}_{galaxy}}$} &
  \multicolumn{1}{c}{\HI{} deficiency} &
  \multicolumn{1}{c}{M$_{\textrm{\HI{}}_{tail}}$} &
  \multicolumn{1}{c}{$\frac{M\HI{}_{tail}}{M\HI{}_{miss}}$} \\  
  \multicolumn{1}{c}{} &
  \multicolumn{1}{c}{[\Msolar{}]} &
  \multicolumn{1}{c}{} &
  \multicolumn{1}{c}{[\Msolar{}]} &
  \multicolumn{1}{c}{} \\  
\hline
  VCC 1249 & $<$\,8.2E6 & $\infty$ & 5.1E7 & 0.20\\
  VCC 865 & 9.9E8 & -0.05 & 7.0E7 & \\
  VCC 497 & 1.5E9 & 0.63 & 9.0E7 & 0.02\\
  VCC 465 & 1.6E9 & 0.20 & 2.0E8 & 0.22\\
  VCC 1516 & 1.2E9 & 0.35 & 2.6E8 & 0.17\\
  VCC 630 & 4.1E8 & 1.09 & 4.0E7 & 0.01\\
  VCC 2066 & 2.0E8 & 0.88 & 9.0E8 & 0.69\\
  VCC 979 & 1.9E8 & 1.26 & 5.4E7 & 0.01\\
  VCC 1987 & 3.4E9 & 0.05 & 4.1E8 & 1.03\\
  VCC 307 & 5.5E9 & 0.00 & 1.8E8 & \\
  HI1225+01 & 1.6E9 & -0.31 & 8.3E8 & \\
  VCC 836 & 4.1E8 & 0.80 & 3.8E8 & 0.17\\
\hline
\end{tabular}
\end{table}

Of the twelve galaxies in table \ref{tab:virgoggalas}, seven are significantly deficient while the others are non-deficient. The lack of correlation between deficiency and presence of a stream is strengthened given that the vast majority of deficient galaxies in the cluster have no reported streams.

Table \ref{tab:virgoggalas} also gives the ratio of the gas detected within a tail compared to the amount of  gas lost according to the deficiency parameter (M\HI{}$_{tail}$/M\HI{}$_{miss}$ column). The majority of tails are simple, linear, one-sided features, which contain much less than the amount of missing gas of their parent galaxies. This suggests that after gas removal a rapid phase change or dispersal of the \HI{} is necessary and sufficient to explain \textit{most} of the features. This is a simple, appealing view of the evolution of the streams but there are two serious caveats. Firstly, as discussed in T16, there are a small number of cases where the mass in the stream appears excessively large in relation to the parent galaxy. Secondly, it is unclear whether such a process is compatible with the presence of a few very long, massive streams~: are those particular features somehow prevented from dissipating~? We therefore need a way to predict how many galaxies should be actively losing gas and the conditions which can render the gas undetectable.

\section{How many streams do we expect to find ?}
\label{sec:predictedstreams}
There are three aspects to the problem : whether streams are observationally detectable, how many are currently forming, and the evolution of the streams as they disperse into the ICM. In this section we examine the first two aspects of observational limitations and stream formation rate. We will examine their evolution in section \ref{sec:streamgobyebye}.

\subsection{Observational restrictions}
\label{sec:geometry}
Almost all of the shorter streams in table \ref{tab:virgogas} are comparable in size to the Arecibo beam, or longer, and thus should be distinguishable from their progenitor galaxies even with the low-resolution Arecibo surveys. Furthermore they are mostly quite massive, $\sim$\,10$^{8}$\,\Msolar{}. ALFALFA and AGES are both ostensibly far more sensitive than the VIVA survey : AGES reports a 1$\sigma$ column density sensitivity limit of approximately N$_{\rm HI}$\,=\,1.5$\times$10$^{17}$\,cm$^{-2}$ (at a line width of a single 10 \kms{} channel, \citealt{olivia}); ALFALFA is about 5.0$\times$10$^{17}$\,cm$^{-2}$ (\citealt{gm33}); VIVA is almost two orders of magnitude worse, at around 1.0$\times$10$^{19}$\,cm$^{-2}$ (\citealt{chatlas}).

An important caveat is that column density is not necessarily a good sensitivity indicator. What the observations are actually sensitive to is the total mass within the beam. Gas can have an arbitrarily high value of N$_{\rm HI}$, but if its mass is too small then the observations may not detect it (e.g. a dense but low mass feature would have a low filling factor within a large telescope beam). Counter-intuitively, a survey with a smaller beam and worse N$_{\rm HI}$ sensitivity may actually be more suitable for detecting low-mass features, provided their N$_{\rm HI}$ is sufficient. This `beam dilution' is particularly important for single-dish telescopes, and interferometers are better adapted to detecting structures smaller than the beam scale.

With this in mind, we can compute the detectability of a stream if we make the idealised assumption that the stream is a linear, uniform-density cylinder. In this case the signal to noise ratio (S/N) from a given survey will depend on the mass, length, velocity profile, and orientation of the stream with respect to the observer. Orientation has two effects, firstly on the projected length $L_{p}$~:
\begin{equation}
L_{p} = L\,sin(i)
\label{eqt:beamlength}
\end{equation} 
Where $L$ is the intrinsic length of the stream and $i$ is the inclination angle to the line of sight, such that $i$$\,$=$\!$0.0$^{\circ}$ if line of sight is parallel to the longest axis of the stream and 90$^{\circ}$ if perpendicular. The number of beams the stream spans in projection will be given simply by $L_{p}/B$ where $B$ is the beam size. The stream can never (by definition) be smaller than a point source in any survey - it must always appear to span at least one beam. The form of the stream in observational data will be its true shape convolved with the telescope beam, but a reasonable approximation is given by :
\begin{equation}
N_{beams} = max\Big( 1.0,\,\frac{L\,sin(i)}{B} \Big)
\label{eqt:nbeams}
\end{equation} 

Secondly, orientation has a very similar effect on how many velocity channels the stream spans. Recall that the relevant parameter for detectability is the mass contained in each beam in each channel :
\begin{equation}
M = \frac{M_{total}}{N_{beams}\,N_{chans}}
\label{eqt:mnice}
\end{equation}
Where $M_{total}$ is the total mass of the entire stream. Given the standard equation for converting \HI{} flux to mass :
\begin{equation}
M_{\rm{HI}} = 2.36\times10^{5}\,d^{2}\,F_{\rm{HI}}
\label{HIMass}
\end{equation}
Where for $M_{\rm{HI}}$ in solar units, $F_{\rm{HI}}$ is the total flux in Jy\,\kms{}, which for a single channel is given by $S/N\,\sigma_{rms}\,w$, where $\sigma_{rms}$ is the $rms$ noise level of the survey, $w$ is the velocity width of the channel (in \kms{}), and $d$ is the distance in Mpc. We can combine the above equations to calculate the S/N of a stream based on its intrinsic parameters and the survey capabilities~: 
\begin{equation}
S/N = \frac{M_{total}}{N_{beams}\,N_{chans}\,2.36\times10^{5}\,d^{2}\,\sigma_{rms}\,w}
\label{eq:streamSN}
\end{equation}

If we disregard the cases where the stream is contained within a single beam or channel, then by equations \ref{eqt:nbeams} and its velocity counterpart (we assume a velocity gradient of magnitude $V$ along the longest axis of the stream), we can re-write equation \ref{eq:streamSN}~:
\begin{equation}
S/N = \frac{M_{total}}{(\sfrac{L}{B})\,sin(i)\,V\,cos(i)\,2.36\times10^{5}\,d^{2}\,\sigma_{rms}}
\label{eq:streamSNnotruncation}
\end{equation}
Thus, while the noise level of the survey remains critical, the survey resolution only determines where the S/N is truncated (see below) but has no other influence over the curve. This may seem counter-intuitive : for example, if one smoothes data in velocity, the flux is spread into fewer velocity channels and so the S/N increases. Of course this procedure also improves the $\sigma_{rms}$, and in practice the resolution and $\sigma_{rms}$ are not truly independent : to get the same $\sigma_{rms}$ when the velocity resolution is higher requires a longer integration time.

\begin{figure}
\centering
\includegraphics[width=85mm]{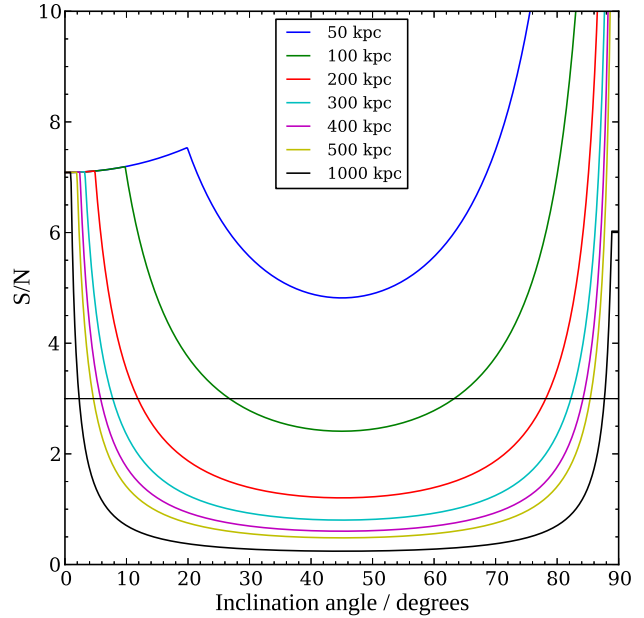}
\caption{Expected S/N level of a linear stream of mass $M_{HI}$ = 1.45$\times$10$^{8}$\Msolar{} and velocity width 500\,\kms{} as a function of viewing angle, for various lengths, assuming survey capabilities equal to AGES (beam size of 17 kpc, $rms$ of 0.6 mJy, and channel width 10 \kms{}). The black line shows a constant S/N level of 3.0, for reference. The vertical axis has been truncated, with an actual peak S/N $>$ 100.0. The `cut' at low inclination angles occurs when the stream spans less than 1 beam in projected length - a similar cut occurs at high angles when the stream spans less than 1 velocity channel.}
\label{fig:streamsAASN}
\end{figure}

Using equation \ref{eq:streamSNnotruncation}, we plot how S/N varies with inclination angle for a linear stream of given parameters (figure \ref{fig:streamsAASN}). At low angles, flux is projected into a short spatial length, giving a high S/N despite a wide spread in velocity. At intermediate angles the flux is spread out both in velocity and space, minimizing the S/N. At high angles, the S/N increases - although it is now highly spatially extended, its projected velocity width is very small. For any given stream, there is a range of inclination angles within which it can be detected.
 
Perhaps more usefully, we can also consider the streams as a population. If we assume the streams have a random orientation with respect to the observer, equation \ref{eq:streamSNnotruncation} can be used to find the range of inclination angles at which a stream of any given parameters will be detectable - so giving the detectable fraction. The projected length of a stream should be at least two beams, otherwise they will not be distinguishable from their parent galaxies\footnote{\cite{greatscott} discuss a search for spectral asymmetries. The effect is generally quite subtle, but may prove fruitful for a future examination of the AGES and/or ALFALFA Virgo data.}. This means their detectable fraction never reaches 100\%.

We do not know the properties of the entire population of streams in Virgo, but we do know about those which have been detected in VIVA. Assuming these are representative of the true stream population, then by this method we can determine how many such features should be detectable to the ALFALFA and AGES surveys. Their median mass (see table \ref{tab:virgogas}) is 1.45$\times$10$^{8}$\Msolar{}. We estimate their intrinsic lengths and velocity widths from the maximum observed values, 60 kpc and 110 \kms{} respectively. The expected detection fraction is rather high, around 70\% for both AGES and ALFALFA~: projected length is the limiting factor in this regime, not total mass, hence the identical spatial resolution of the surveys gives identical detection fractions.

The expected number of detected streams in any given survey depends on 1) the number of \HI{}-detected galaxies in the survey region (439 for ALFALFA in the VCC region, 105 for AGES); 2) what fraction of those galaxies actually have streams, which we take to be 15\% based on VIVA; 3) the geometrical correction for how many streams that exist should also be detectable, i.e. 70\%. This gives expected stream numbers of 46 for ALFALFA and 11 for AGES. The actual numbers are 5 for ALFALFA and 2 for AGES. Hence the geometric correction is insufficient to explain their low detected numbers. We now consider which galaxies are expected to be currently producing streams in the first place.

\subsection{Stream formation}
\label{sec:koppen}
The modelling of \cite{koppen} provides an analytic model of ram pressure stripping. This considers how much pressure is required to strip the \HI{} down to its observed radius (P$_{req}$) in comparison with an estimate of the local pressure the galaxy is actually experiencing (P$_{loc}$), given its position in the cluster. Given the uncertainties, a galaxy may be actively stripping if the ratio P$_{loc}$/P$_{req}$$\,$$\geq$\,0.5. The main advantage to this is that it describes current stripping activity - potentially a much better proxy for the presence of a stream than \HI{} deficiency. The disadvantage is that the necessary data are only available for a small fraction of the galaxies, so we cannot use it to predict the total number of expected streams in this region.

Despite this, the model can make more qualitative predictions. We do not expect every galaxy which is predicted to be an active stripper to have an \HI{} tail, due to the distance uncertainty (which affects the calculated P$_{req}$ value) and the possible geometrical dilution of the tail described in section \ref{sec:geometry}. If, however, ram pressure stripping is indeed the dominant gas-loss mechanism in the cluster, then we expect tails to be more common among galaxies with higher P$_{loc}$/P$_{req}$ ratios. We also expect every galaxy (with only rare exceptions due to harassment and ICM density variations) which has an \HI{} tail to be an active stripper according to the model. We will return to this in section \ref{sec:predstrip}.

\section{Searching for streams}
\label{sec:searching}
\subsection{Data processing}
\label{sec:dprocess}
We test the models of section \ref{sec:predictedstreams} using the two AGES Virgo data cubes described in \cite{me12} and \cite{me13}. The much larger ALFALFA data set is not publically available, and the AGES data sets, though smaller, have the advantage of higher sensitivity. Both cubes are available via the AGES website at the following URL~: \href{http://www.naic.edu/$\sim$ages/}{http://www.naic.edu/$\sim$ages/}.

There are two significant improvements to the data processing algorithms developed since the original analysis. The first is an implementation of the spatial bandpass processing algorithm MEDMED, described in \cite{put}. AGES is a R.A.-parallel drift scan survey, with the baseline level of the spatial bandpass being nominally estimated as the median level of the entire scan. This is adequate for most scans in which galaxies occupy only a few percent of the bandpass, but where bright, extended sources are present, the baseline average value is over-estimated. This results in `shadows' in the cube in R.A. (see \citealt{m10}). As in \cite{me14}, we use a Python-based version of MEDMED that splits the scan into five boxes, measures the median of each, and then uses the median of the medians as the baseline. This almost completely eliminates the shadows which would otherwise hamper the search for extended emission in the affected areas.

\begin{figure}
\begin{center}  
  \subfloat[Original cube]{\includegraphics[width=85mm]{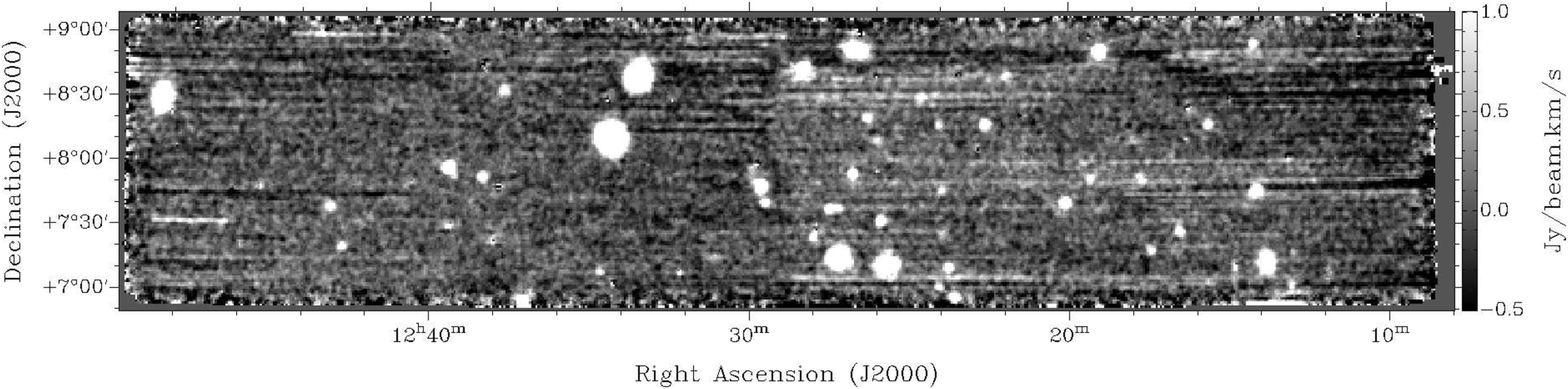}}\\
  \subfloat[After cleaning]{\includegraphics[width=85mm,]{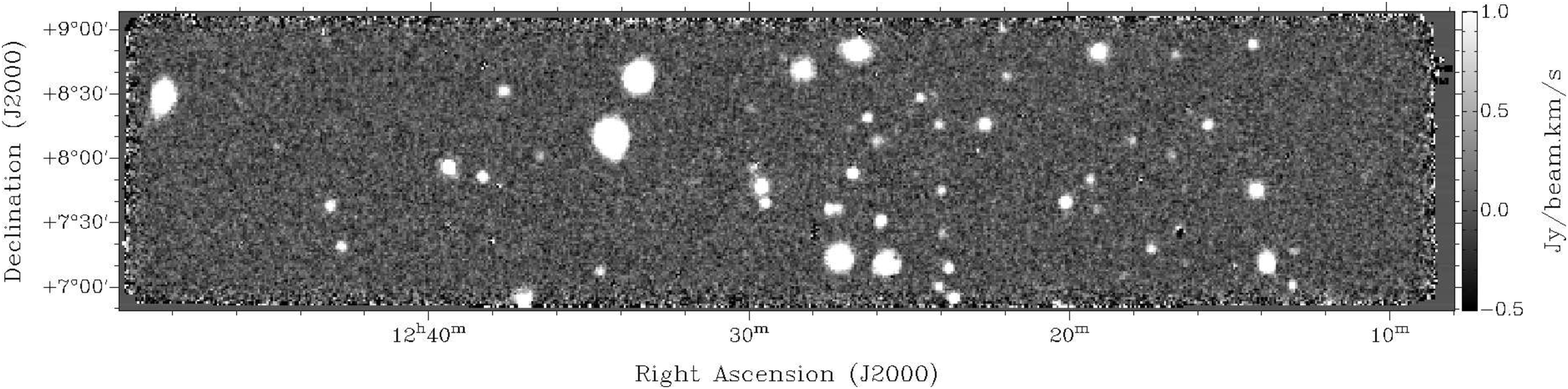}} 
\caption{Moment 0 (integrated flux) maps of the AGES VC1 data cube, using the same velocity range (100-3,000 \kms{}) and with the same colour scale in both cases. The upper figure uses the standard data cube which has no additional processing besides hanning smoothing; the lower figure uses the cleaning techniques described in section \ref{sec:searching}. The \textit{rms} of the spatial bandpass in the cleaned image is approximately 33\% lower than in the raw image, whereas the mean pixel value is a factor of three lower in the cleaned image.}
\label{fig:cleaning}
\end{center}
\end{figure}

The second change, also described in \cite{me14}, is to fit a second-order polynomial to the spectrum along each pixel in the cube. While the $rms$ of each spectrum is not affected, the removal of the baseline variation from pixel to pixel greatly improves the `cleanliness' of the data, making it much easier to search for extended emission and improving the accuracy of flux measurements on extended structures. The combined effect of these cleaning processes is shown in figure \ref{fig:cleaning}.

\subsection{Search technique}
\label{sec:search}
Another key development in our search for streams is the FITS viewer \textsc{frelled}, described in \cite{me15}. The main benefit here is the user can interactively create moment maps and contour plots, i.e. to find the most appropriate velocity and spatial range with which to examine each galaxy. We can also examine the data in 3D rather than conventional 2D slices, which can make visual identification of extended features much easier.

\begin{figure}
\centering
\includegraphics[width=85mm]{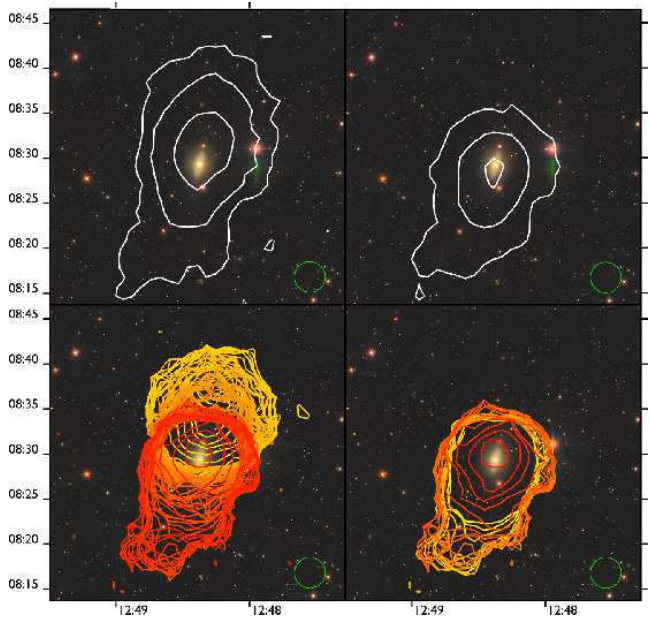}
\caption{Different 2D display techniques for the same galaxy, VCC 2070. The background image is the SDSS RGB image from the standard pipeline, the same in all panels. The top panels show moment 0 contours while the lower panels show renzograms. The left side images use the full velocity range of the galaxy while those on the right are restricted to the velocity range we identify as containing the associated stream. The green circle is the size of the Arecibo FWHM, 3.5$^{\prime}$.}
\label{fig:displayproblems}
\end{figure}

We found that the best way to detect extended emission was through inspection of renzograms (contour maps in which each velocity channel is represented by a different colour, see \citealt{rupp}) for non-circular features. Integrated flux maps, though of greater sensitivity to diffuse gas, tend to be problematic. The galaxies themselves often have marginally resolved gas discs - while the discs tend to have circular \HI{} contours in every channel, their centre is not quite at the same pixel position in each channel. Thus, integrating over the whole velocity range produces non-circular features which are not related to genuine extensions - see figure \ref{fig:displayproblems}.

\begin{figure}
\centering
\includegraphics[width=85mm]{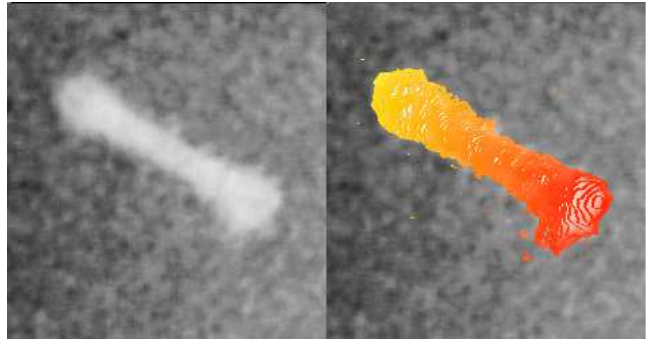}
\caption{Example of using different visualisation techniques to search for extended features. The left panel shows a volume rendering (integrating the flux along the line of sight) of VCC 2070 at an arbitrary angle. The extended gas tail is barely visible. The right panel shows the data cube rendered in exactly the same way, but with a renzogram at 4$\sigma$ overlaid. The extension at the low-velocity tip of the galaxy (right side of the image) is now much more obvious.}
\label{fig:renzoexample}
\end{figure}

Viewing renzograms in 3D shows essentially \textit{isosurfaces} that display constant flux levels as a 3D surface rather than a 2D contour. This gives a powerful advantage in the search for non-circular extensions~: since most channels tend to have circular contours, those with non-circular extended features easily stand out (especially if those features are coherent over several channels). Unlike volume renders, isosurfaces have the  benefit of displaying the flux at an objective level which does not depend on viewing angle (see \citealt{me15} for a full discussion), and the 3D display makes it easier to see which channels possess extensions than the case of 2D renzograms (where the superposition of many channels can be confusing). We show an example in figure \ref{fig:renzoexample}. 

We limited our search to the velocity range 100\,$<$\,$v_{hel}$\,$<$ 3,000 \kms{}, avoiding the Milky Way and high velocity clouds. We constructed renzograms/isosurfaces for every catalogued \HI{} detection in this region, 106 out of our total of 108 Virgo detections for VC1 and VC2 combined. The two omissions were at such low redshifts that Milky Way contamination would make distinguishing any extended emission extremely difficult.

Our procedure was to begin with renzograms at 3$\sigma$ and then increase the S/N level as appropriate. For the brightest sources, extensions are not visible at 3-5$\sigma$ simply because the disc emission is very bright, ``smearing'' the emission into many pixels. Our requirement for a detection was that the non-circular features should be visible at a defined level per volumetric pixel (voxel) across a connected span of at least one beam (in addition to the galaxy's disc) and over at least 3 channels. We catalogued possible streams by the noise level of the connected voxels as either certain ($>$ 6$\sigma$ per voxel), probable ($>$4-6$\sigma$ per voxel) or possible ($>$ 3-4$\sigma$ per voxel). We discuss these levels in detail in section \ref{sec:fdr}.

\begin{figure*}
\centering
\includegraphics[width=170mm]{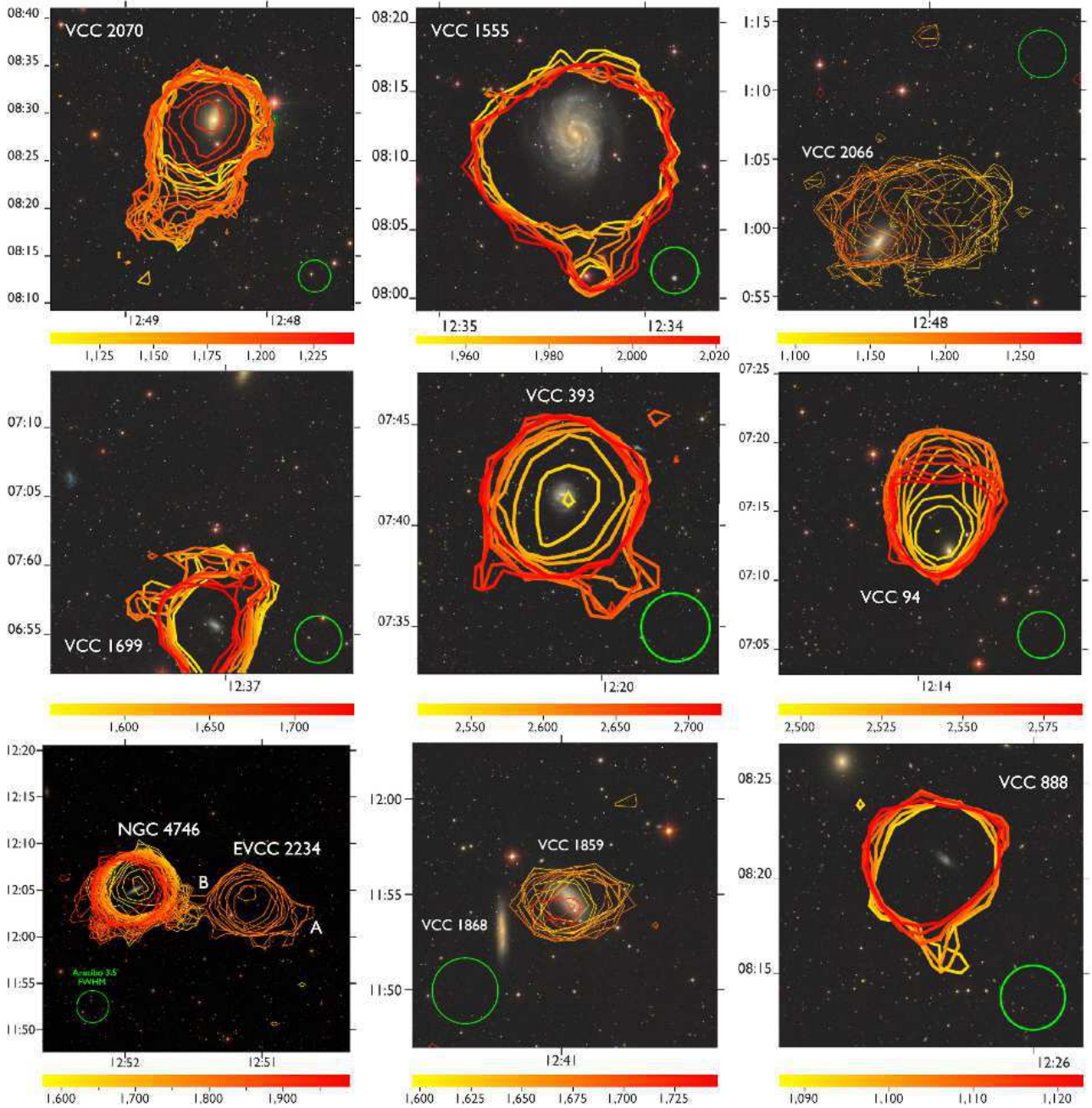}
\caption{Renzograms of the `certain' (VCC 2070, VCC 1555, and VCC 2066 on the top row; EVCC 2234 is in the bottom left panel) and `probable' stream detections. The contour levels are of fixed value with colour indicating the channel. The S/N level of the contour (typically 4\,$\sigma$, equivalent to a column density of $N_{HI}$\,=\,6$\times$10$^{17}$\,cm$^{-2}$) and velocity range of the renzogram (shown by the colour bars in units of \kms{}) have been manually adjusted in each case to reveal the streams most clearly so do not always show the full velocity range of each galaxy - for exact individual values, consult appendix \ref{sec:appendix}. The green circles show the Arecibo beam size.}
\label{fig:goodstreams}
\end{figure*}

Maps of the individual certain or probable detections are shown in figure \ref{fig:goodstreams}. We do not discuss the less confident detections here but present them in the online appendix. The full catalogue of the stream status of all the galaxies in this region (excluding those with no streams of any kind) is shown in table \ref{tab:allvc1streams}.

\begin{table*}
\centering
\caption{Catalogue of all the extended \HI{} features detected in AGES in the Virgo cluster. All measurements are derived from AGES data except for those of AGESVC2 020 (VCC 2066) - as this is on the southern limit of the data where measurements may not be accurate, we use the values from \cite{duc07} instead. The sections divide the table according to stream type, where 0-2 denote possible streams (0 means the detection is certain, 1 probable, and 2 possible), and 3 indicates that the galaxy has noisy contours with no preferred direction for the extensions. For completeness we also show category 4, meaning no detected streams, for those objects where the pressure ratio described in section \ref{sec:koppen} was calculated.}
\label{tab:allvc1streams}
\footnotesize
\begin{tabular}{c c c c c c c c c c}
\hline
  \multicolumn{1}{c}{AGES ID} &
  \multicolumn{1}{c}{Galaxy ID} &
  \multicolumn{1}{c}{Galaxy M\HI{}} &
  \multicolumn{1}{c}{Missing M\HI{}} &
  \multicolumn{1}{c}{Stream} &
  \multicolumn{1}{c}{$\frac{P_{loc}}{P_{req}}$} &
  \multicolumn{1}{c}{Stream length} &
  \multicolumn{1}{c}{Stream M\HI{}} &
  \multicolumn{1}{c}{ $\frac{M\HI{}_{stream}}{M\HI{}_{miss}}$ } \\
  \multicolumn{1}{c}{} &
  \multicolumn{1}{c}{} &
  \multicolumn{1}{c}{[\Msolar{}]} &
  \multicolumn{1}{c}{[\Msolar{}]} &
  \multicolumn{1}{c}{code} &
  \multicolumn{1}{c}{} &
  \multicolumn{1}{c}{[kpc]} &
  \multicolumn{1}{c}{[\Msolar{}]} &
  \multicolumn{1}{c}{} \\
\hline
  AGESVC1 269 & VCC 1555 / NGC 4535 & 1.1E9 & 3.0E9 & 0 & 0.87 & 60 & 3.4E8 & 0.113\\ 
  AGESVC1 270 & VCC 2070 / NGC 4698 & 1.2E9 & 4.3E9 & 0 & 0.21 & 75 & 2.1E8 & 0.049\\
  AGESVC2 025 & EVCC 2234 & 1.0E8 & 1.7E8 & 0 &  & 34 & 5.4E6 & 0.032\\
  AGESVC2 020 & VCC 2066 / NGC 4694 & 2.5E8 & 1.3E9 & 0 & 0.17 & 50 & 9.0E8 & 0.692\\
\hline
  AGESVC1 256 & VCC 1699 & 5.3E8 & 6.0E7 & 1 & 2.33 & 32 & 3.4E7 & 0.567\\
  AGESVC1 202 & VCC 94 / NGC 4191 & 1.7E9 & -3.7E8 & 1 &  & 98 & 1.5E8 & -0.405\\
  AGESVC1 246 & VCC 888 & 2.4E8 & 7.7E8 & 1 &  & 35 & 2.9E6 & 0.004\\
  AGESVC1 210 & VCC 393 / NGC 4726 & 5.3E8 & 1.4E9 & 1 &  & 65 & 2.0E7 & 0.014\\
  AGESVC2 022 & VCC 1972 / NGC 4647 & 4.1E8 & 6.3E8 & 1 & 0.30 & 51 & 1.7E7 & 0.027 &\\
  AGESVC2 033 & VCC 1859 / NGC 4606 & 4.5E7 & 2.8E9 & 1 & 0.01 & 17 & 5.2E7 & 0.002\\
\hline
  AGESVC1 229 & VCC 667 & 2.9E8 & 2.7E9 & 2 &  & 42 & 3.0E7 & 0.011\\
  AGESVC1 204 & None & 3.6E8 & 3.5E8 & 2 &  & 46 & 1.6E7 & 0.046\\
  AGESVC1 207 & VCC 199 / NGC 4224 & 8.3E8 & 4.1E9 & 2 &  & 49 & 3.0E7 & 0.007\\
  AGESVC1 232 & None & 8.2E7 & 1.0E8 & 2 &  & 33 & 2.5E6 & 0.025\\
  AGESVC1 278 & VCC 1011 & 1.7E8 & 7.1E8 & 2 & 0.23 & 37 & 3.0E6 & 0.004\\
  AGESVC1 238 & VCC 688 / NGC 4353 & 1.5E8 & 1.3E9 & 2 &  & 23 & 1.2E7 & 0.009\\
  AGESVC1 284 & VCC 740 & 1.9E8 & 3.8E8 & 2 &  & 42 & 1.1E7 & 0.029\\
  AGESVC1 248 & VCC 713 / NGC 4356 & 8.9E7 & 6.5E9 & 2 &  & 66 & 3.9E7 & 0.006\\
  AGESVC1 272 & VCC 1725 & 1.3E8 & 5.0E8 & 2 & 0.40 & 36 & 4.3E6 & 0.009\\
  GLADOS 001 & None & 6.5E7 & 2.1E8 & 2 &  & 46 & 1.1E7 & 0.052\\
  AGESVC1 286 & VCC 514 & 1.2E8 & 1.9E9 & 2 &  & 40 & 6.5E6 & 0.003\\
  AGESVC1 235 & None & 2.6E7 & 7.1E7 & 2 &  & 59 & 5.7E6 & 0.080\\
  AGESVC2 063 & None & 1.5E7 & 4.5E7 & 2 &  & 17 & 1.0E6 & 0.022\\
  AGESVC2 018 & VCC 1868 / NGC 4607 & 2.5E8 & 1.0E9 & 2 & 0.25 & 40 & 1.9E7 & 0.019\\
  AGESVC2 027 & NGC 4746 & 8.3E8 & 1.8E9 & 2 &  & 34 & 5.8E7 & 0.032\\
  AGESVC2 025B & EVCC 2234 & 1.0E8 & 1.7E8 & 2 &  & 34 & 8.1E6 & 0.048\\
\hline  
  AGESVC1 259 & VCC1952 & 1.4E8 & 3.8E8 & 3 &  & 36 & \\
  AGESVC1 212 & VCC1205 / NGC 4470 & 4.7E8 & 9.4E8 & 3 & 0.52 & 39 & \\
  AGESVC1 261 & VCC1394 & 2.5E7 & & 3 &  & 32 & \\
  AGESVC1 225 & VCC1791 & 4.5E8 & 3.7E8 & 3 &  & 59 & \\
  AGESVC1 281 & VCC1249 & 5.1E7 &  & 3 &  & 40 & \\
  AGESVC1 244 & VCC566 & 4.1E8 & -8.5E7 & 3 &  & 47 & \\  
  AGESVC1 220 & VCC318 & 2.1E9 & 6.1E8 & 3 &  & 81 & \\  
  AGESVC1 216 & VCC207 & 1.5E8 & 2.0E8 & 3 &  & 34 & \\ 
\hline
  AGESVC1 268 & VCC 2007 & 2.5E7 & 1.3E8 & 4 & 0.02 & & & & \\
  AGESVC1 240 & VCC 938 / NGC 4416 & 2.6E8 & 6.5E8 & 4 & 0.05 & & & & \\
  AGESVC1 292 & VCC 1575 & 9.0E7 & 6.9E8 & 4 & 0.06 & & & & \\
  AGESVC1 279 & VCC 1193 / NGC 4466 & 1.7E8 & 1.5E8 & 4 & 0.10 & & & & \\
  AGESVC1 263 & VCC 1758 & 2.0E8 & 3.9E8 & 4 & 0.16 & & & & \\ 
  AGESVC2 019 & VCC 1955 / NGC 4641 & 3.7E7 & 4.0E8 & 4 & 0.18 & & & & \\    
 \hline 
\end{tabular}
\end{table*}

We catalogued four streams as certain, six as probable, and sixteen as possible. We also found eight galaxies with \HI{} contours with no indications of asymmetry but which had a distinctly `noisy' appearance, sometimes even at the 7$\sigma$ level. None of the remaining 73 galaxies showed any indications of any unusual \HI{} features, though 20 of these were of rather low S/N ($<$\,10) so extensions would be difficult to detect.

\subsection{Measuring the streams}
\label{sec:measurements}
Unlike the galaxies in \cite{me14}, the Virgo objects are marginally resolved, and thus cannot be measured as point-sources and subtracted to allow objective measurements. This means we are compelled to resort to more subjective procedures. We use \textsc{FRELLED}'s capability to define volumes of arbitrary shape and sum the flux within them, manually defining volumes we believe only contain flux from the extended \HI{}. Thus the estimates of the stream masses in table \ref{tab:allvc1streams} should be treated with caution (the estimated ratios M\HI{}$_{stream}$/M\HI{}$_{miss}$ are similar to the values for the \citealt{chung} sample as in table \ref{tab:virgoggalas}, though on the low side). We do not attempt this procedure at all for galaxies with noisy contours.

We measure the length of the streams as the distance from the centre of the galaxy to the most distant extension of the stream contours. We do not apply the correction for beam smearing described in \cite{wang} as the difference is only a few kpc for our sources, and our errors are dominated by the problems of determining the edge of the parent galaxy's disc. Additionally, the beam size means that we cannot accurately measure the thickness of the streams - the apparent visual difference in thickness of the streams compared to their parent galaxies simply reflects the relative brightness of stream and galaxy, not their dimensions.

\subsection{Comparisons to other data}
\label{sec:inter}
Six galaxies in our sample have interferometric observations. VCC 2070 and VCC 1555 were observed with VIVA, with shorter extensions detected at similar orientations to those detected here (see appendix A). The tail of VCC 2066/2062 has a very similar overall morphology in both AGES and the VLA observations of \citealt{duc07}, though the VLA data shows structures within the tail that AGES cannot resolve. The morphology of the gas cloud close to VCC 1249 is very similar in AGES (see T12 figure 21) to the KAT7 observations of \cite{sorgho}, hereafter S17 (see their figure A3). In general the AGES data support the existing interferometric observations, in some cases extending the length of the tails significantly.

There are two exceptions. One is VCC 1205 (see section \ref{sec:VCC1205}), the extension of which is not described by S17. However this is not surprising - their observations are nominally 5 times less sensitive than ours, with a 1\,$\sigma$ column density sensitivity of 8$\times$10$^{17}$\,cm$^{-2}$. This is reduced further at the position of VCC 1205 as that region was only observed with KAT7. Additionally, S17 only detect a small part of the ALFALFA Virgo 7 cloud complex described in \cite{kent09} (which similarly happens to be at the edge of the field where only KAT7 data was taken), and the authors attribute this to the gas being at low column density and below the sensitivity of the interferometer.

More puzzlingly, the feature described in S17 associated with AGESVC1 293 is not visible in the AGES data. Here the KAT7 and WSRT pointings overlap, though the galaxy is near the edge of the survey fields where sensitivity is again somewhat reduced (see S17 figure 3). It has an \HI{} mass of 2.0$\times$10$^{7}$\Msolar{} and a W50 of 87\,\kms{}. The morphology of the source is irregular, so beam dilution may play some role, but the main feature is comparable in size to the Arecibo beam. If entirely contained within the Arecibo beam, the average column density in each 10\,kms{} AGES velocity channel would be 1.2$\times$10$^{18}$\,cm$^{-2}$, well above the AGES sensitivity limit. Deeper observations are needed to confirm the existence of this source.

% Notes we may want to add to appendix :
% VCC 2070 - something seen in moment map, nothing in channel maps
% VCC 1555 - something seen in moment map, nothing in channel maps
% VCC 2066 - AGES MHI 2.3E8; Duc 1.1E9. Near edge of cube, could also be due to improper AGES measurement
% VCC 1249 - AGES MHI 5.1E7; Sorgho 7.0E7; AGES W50 29 W20 63 km/s

\subsection{Estimating the false detection rate}
\label{sec:fdr}
Our criteria for ``probable'' detections being at least 4$\sigma$ may seem weak, implying that we could expect a high fraction of our results to be spurious. In this section we examine this statistically by three different methods. Throughout, it is crucial to remember that our criteria for identification relies not just on S/N but also on the spatial and velocity extent of the features.

\subsubsection{The number of similar features in galaxy-free regions of the data}
\label{obsearch}
If the claimed streams are actually just fluctuations in the noise then they should be present throughout the entirety of the data cube. Although difficult to disentangle from bright, marginally resolved galaxies, in empty regions it is straightforward to find and measure such features using objective, repeatable procedures.

\begin{figure}
\centering
\includegraphics[width=84mm]{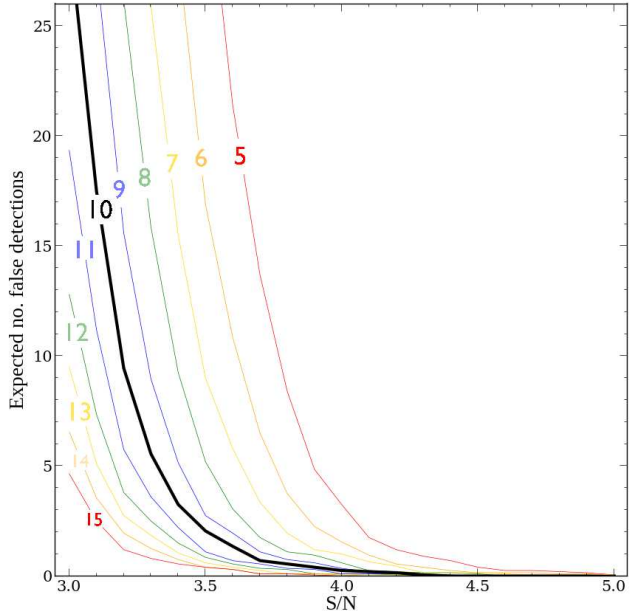}
\caption{Expected number of false detections of streams at different S/N levels (x-axis) and number of connected pixels (annotations). This is based on a search of the empty regions of the the VC1 data cube, described in the text. The y-axis value has been scaled according to the volume spanned by the galaxies and their streams, and truncated at 26 (the total number of certain, probable and possible streams).}
\label{fig:fdr}
\end{figure}

We begin by masking the galaxies and the identified streams, which accounts for about 5\% of the total volume of the VC1 data cube. For the remaining pixels, we use the \textsc{stilts} package (\citealt{stilts}) to match groups of connected pixels at or above a range of S/N levels. We then quantify the number of groups of pixels based on both the number of connected pixels and the S/N level. Given the detection rate in the galaxy-free regions and the total volume searched for extensions, we can estimate the number of false detections we expect around the galaxies. The results are shown in figure \ref{fig:fdr}.

At 4$\sigma$ per voxel, even groups of only five connected pixels (approximately one beam in one channel) are expected to be so rare that they are unlikely to cause significant contamination. Five pixels is extremely small - ten is more reasonable for our ``probable'' and ``certain'' detections, which generally fill at least one beam and are found in three or more channels. With ten or more pixels, streams can be considered reliable even at 3.5$\sigma$. Furthermore, the curves plotted in figure \ref{fig:fdr} will significantly overpredict the number of false detections~: this figure assumes that a spurious clump found at any location within the searched volume would be mistaken for an extension. In reality, the clump would have to appear in a very specific, much smaller region - if it coincided with a galaxy it would not be visible at all, whereas if it was too far from the galaxy it would be identified as an independent object rather than an extension. We conclude that our ``certain'' and ``probable'' identifications suffer a negligible rate of false positives, though doubtless there may be contamination in our ``possible'' features. It should be remembered that only 36\% of the galaxies inspected in the VC1 cube showed any signs of extensions at all - even the most modest potential streams included in our catalogue are notably different from most of the galaxies in the data.

\subsubsection{Searching for simulated streams of known parameters}
\label{sec:quicksearch}
Using artificial galaxies and streams with real noise, we can also test our subjective search techniques. In this way we can examine (a) whether the noise will allow us to detect features as weak as those we claim to have detected and (b) whether we would visually identify more false positives that the objective procedure suggests.

We create an artifical galaxy with parameters (S/N = 55.0 and velocity width of 220 km/s) based on the median values of the galaxies we have identified as having sure or probable streams. For the galaxy we use either a simple point source or a more realistic model based on the radially averaged profile of the real, marginally resolved VCC 975. We create a stream beginning with a grid of 6$\times$2 pixels of uniform S/N level (3, 4 or 5$\sigma$), which we then convolve with a 3.5$^{\prime}$ Gaussian. This we then add to either the first 1, 2, or 3 consecutive channels in the artifical galaxy data set, our aim being to explore the detectability of the weakest features.

\begin{figure}
\centering
\includegraphics[width=84mm]{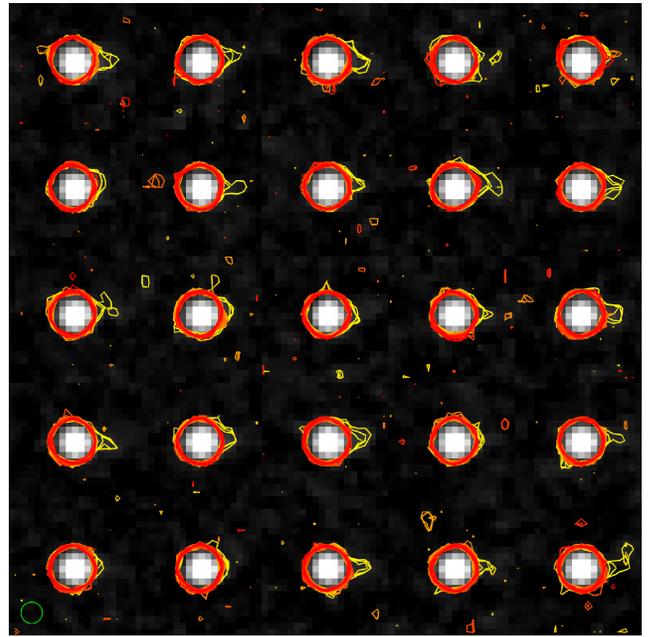}
\caption{Example subset of artifical galaxies with extensions combined with real noise from the AGES VC1 data cube. In this case the extension has an original S/N level of 3$\sigma$ (boosted slightly by the presence of the galaxy) with the contour at 3.5$\sigma$. This approximately corresponds to some of our faintest claimed detections. When visible, the extension is usually clear, whereas there are almost no visible false positives even at this low S/N level.}
\label{fig:fakes}
\end{figure}

Next, we run a Python script that chooses random pixels within the masked VC1 cube and checks for the presence of masked pixels within the appropriate surrounding volume. If any are found, another pixel is chosen and this is repeated until a suitable region is found. We then add the galaxy and stream into this region of pure noise. The data in this region is extracted, and the procedure is repeated 100 times to create galaxies+stream+noise data cubes. The properties of the galaxy and stream do not vary so this procedure tests only the influence of the noise. An example subset of the data is shown in figure \ref{fig:fakes}, plotting renzograms of the point source case, with the stream extending from the centre of the galaxy to the right.

We inspect the final data set visually, varying the contour level from 3.0-5.0$\sigma$ in steps of 0.5. At each level, we record how many galaxies show clear signs of the aritifical stream in their renzograms. We require a detection to span at least one beam, the same number of channels as it was injected in, and be present at the correct location. We also count the number of galaxies with similar features that are not at the correct location, i.e. false positives. Knowing the size and orientation of the stream makes this process very fast, enabling us to explore a large parameter space of detectability (note that the detection criteria deliberately probe very low S/N levels and channel numbers, and do not reproduce the criteria used in the actual search).

\begin{figure}
\begin{center}  
  \subfloat[1 channel]{\includegraphics[width=40mm]{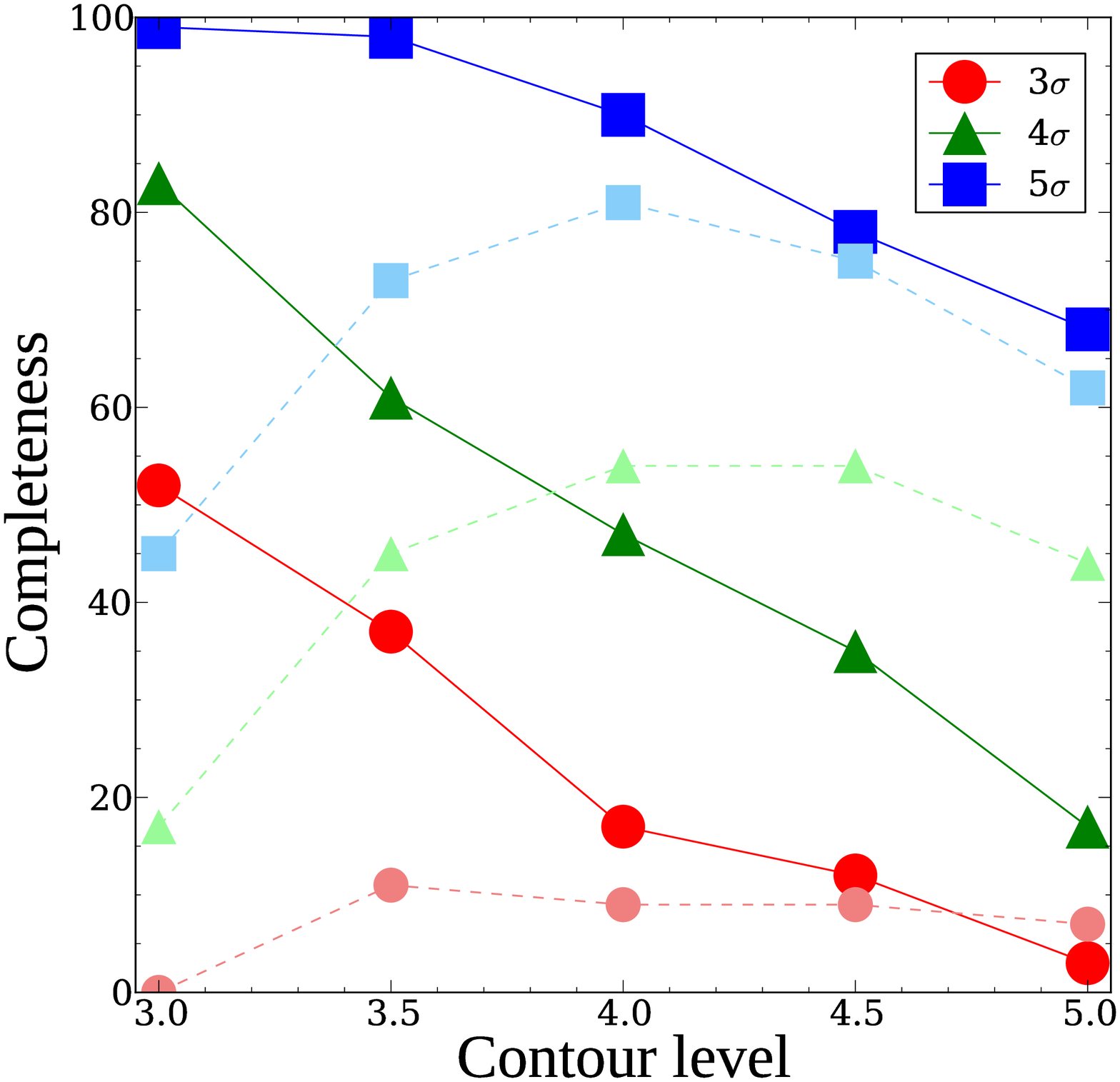}}
  \subfloat[1 channel]{\includegraphics[width=40mm]{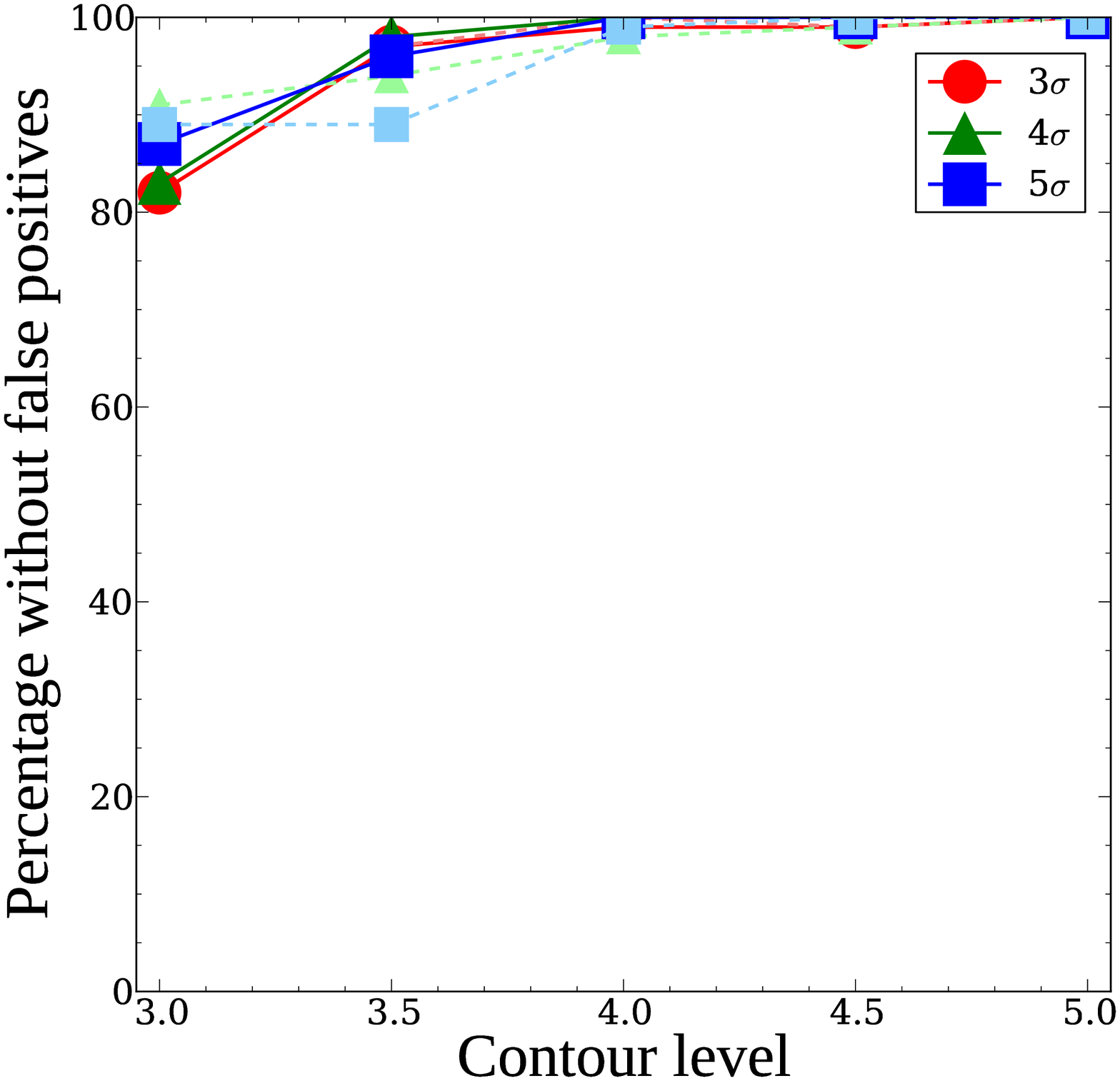}}\\
  \subfloat[2 channels]{\includegraphics[width=40mm]{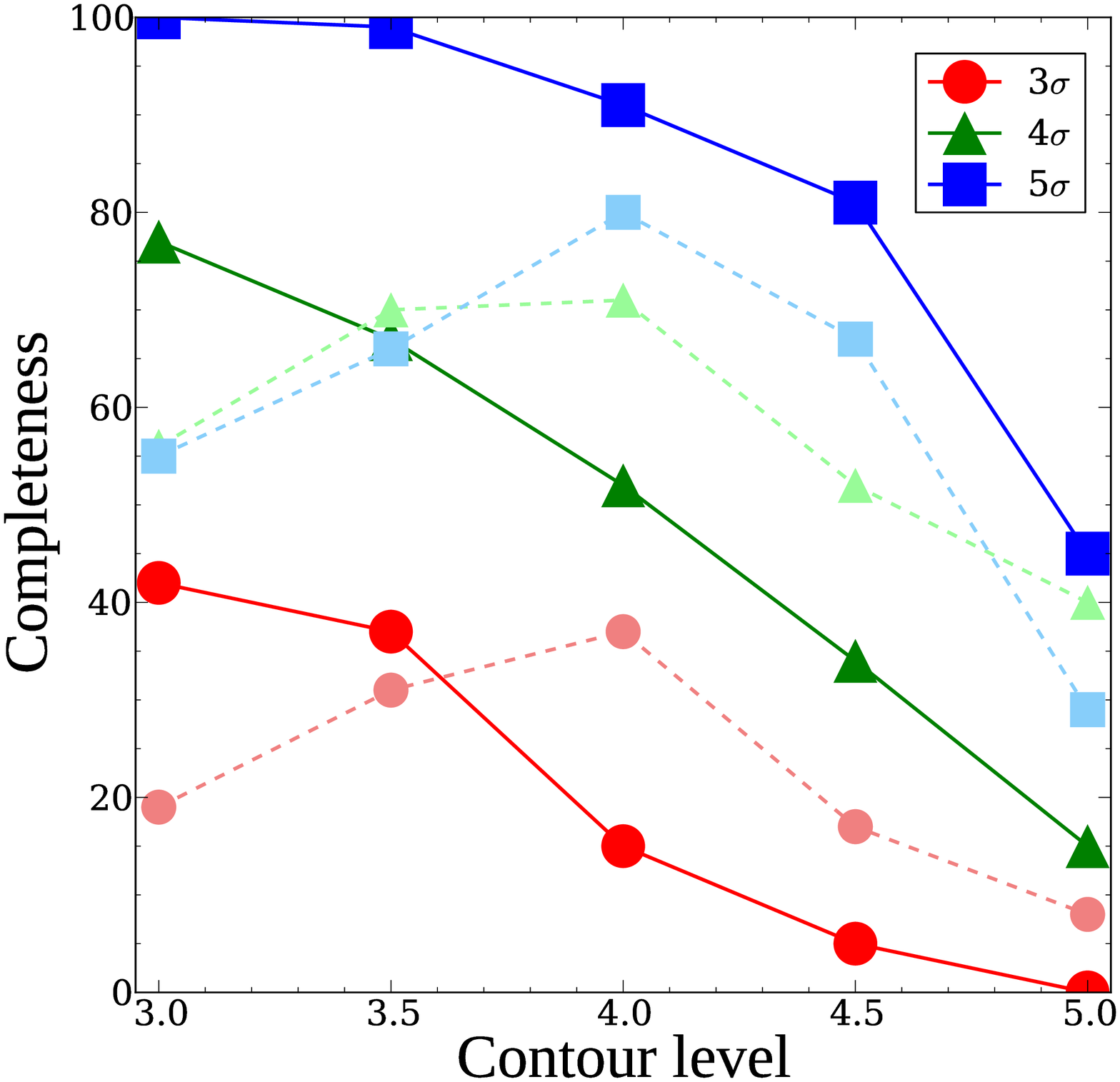}}
  \subfloat[2 channels]{\includegraphics[width=40mm]{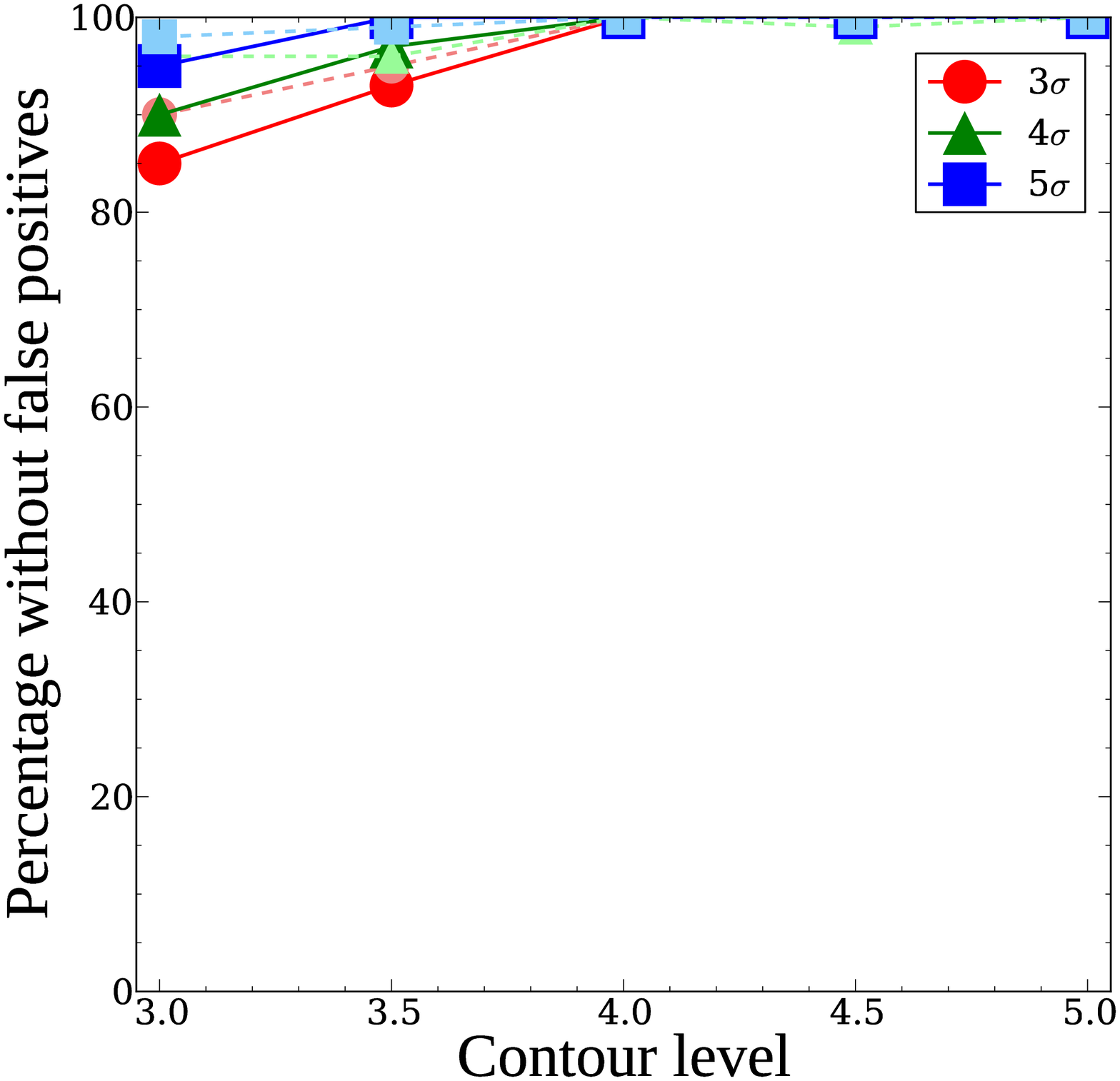}}\\
  \subfloat[3 channels]{\includegraphics[width=40mm]{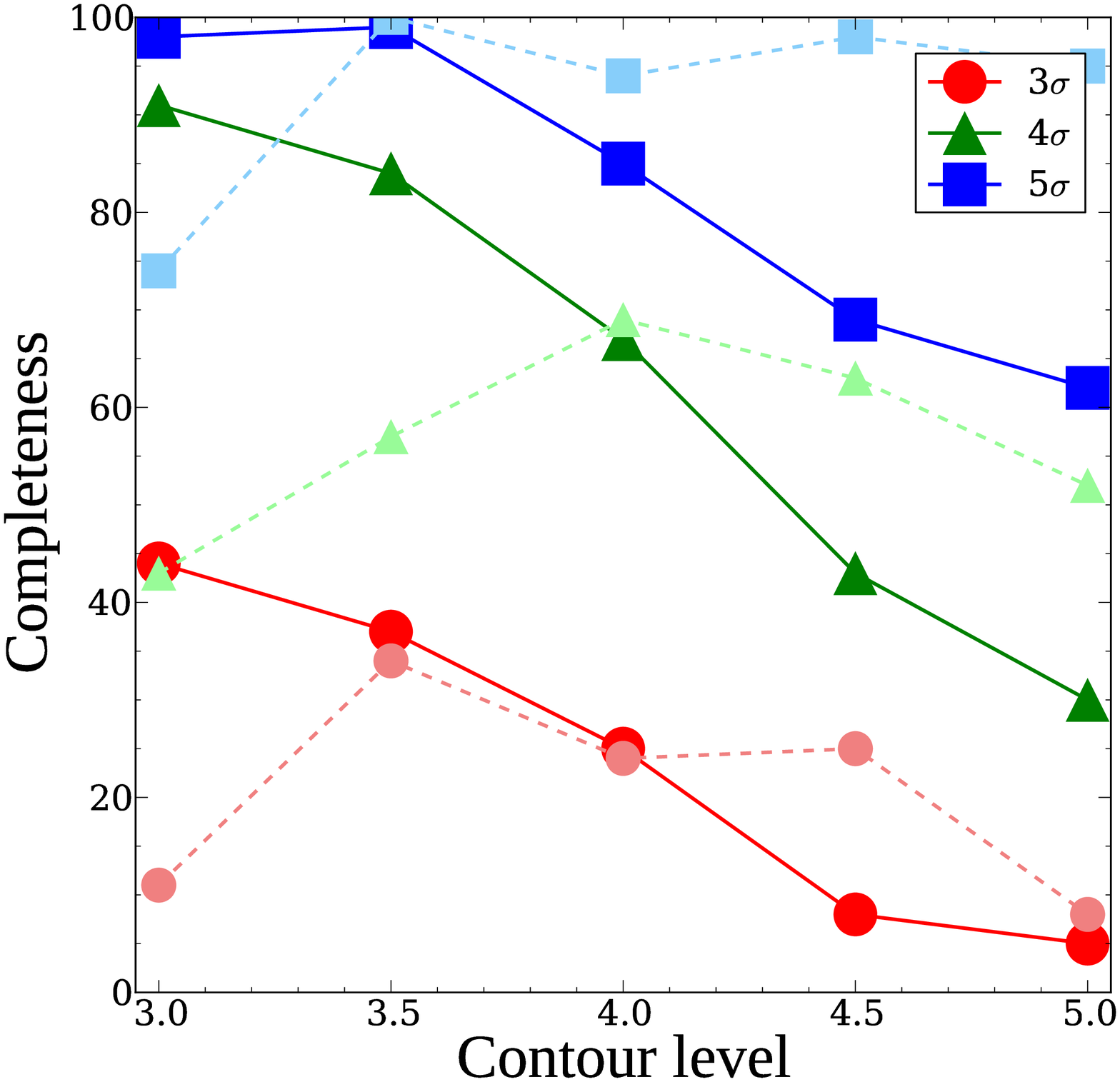}}
  \subfloat[3 channels]{\includegraphics[width=40mm]{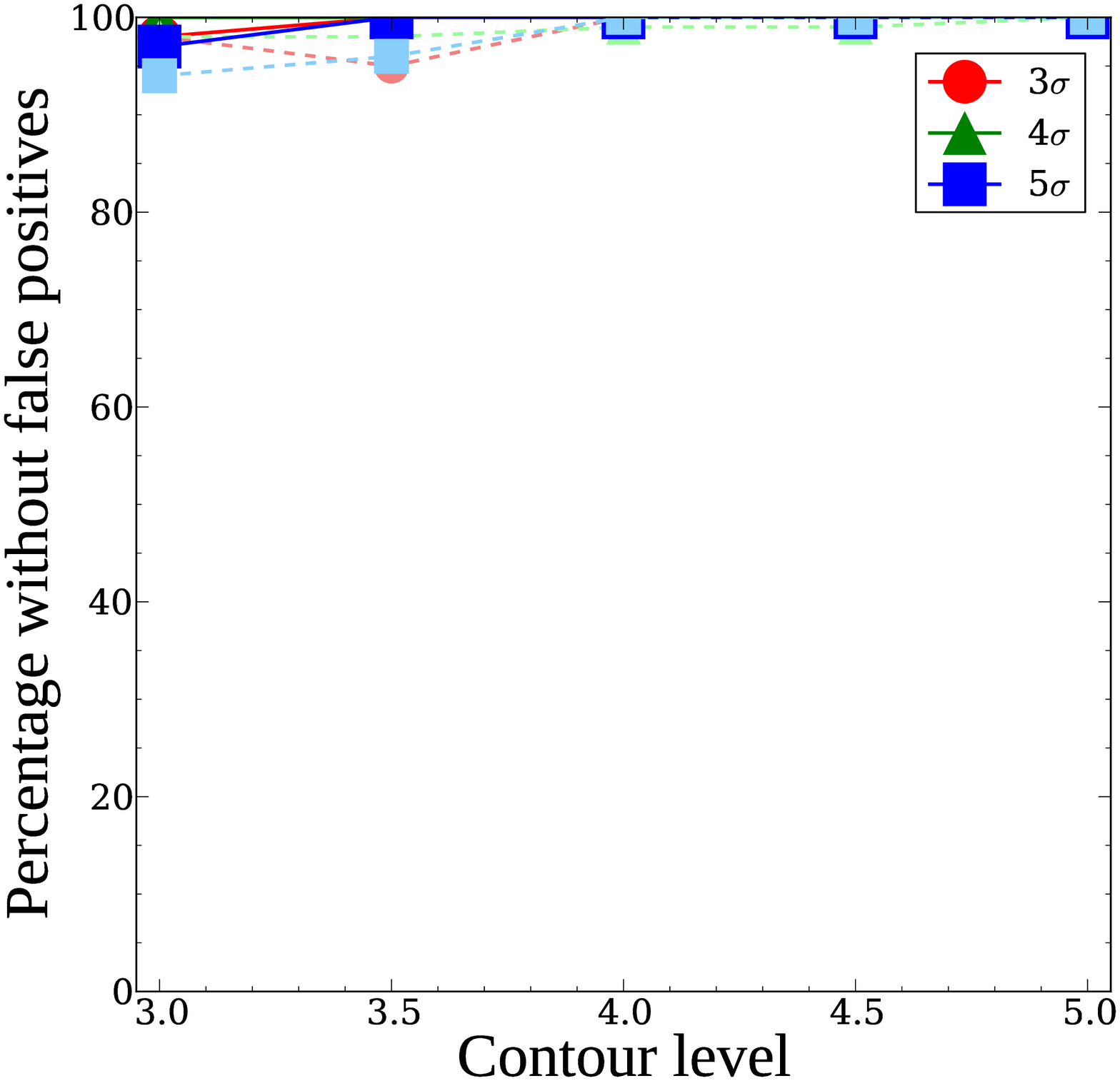}} 
\caption{Completeness levels (left) and false positive rate as a function of contour level for sources injected into different numbers of channels (1, 2, and 3 from top to bottom). Line colour indicates the original S/N level of the injected source. Solid lines show the case of using a point source for the galaxy while dashed lines use a more realistic galaxy profile.}
\label{fig:allthegraphs}
\end{center}
\end{figure}

The results are shown in figure \ref{fig:allthegraphs}. Since the streams are added to a bright galaxy, their final S/N is usually slightly higher than their initial value, hence it is sometimes possible to detect nominally faint streams at surprisingly high S/N values. ``Completeness'' here is in the usual sense, i.e. the fraction of known streams which were detected. We cannot properly measure reliability here, as this depends on the number of injected streams which is a free parameter in this exercise. Instead the independent parameter is the number of false positives.

We find that at very low S/N levels, the data is essentially one contiguous set of pixels from which nothing whatsoever can be discerned. At high S/N levels one sees only the bright galaxies. In between these extremes, there does not exist a realm in which one sees significant numbers of features resembling the streams we see around actual galaxies. Such features do exist, but are always rare, and hence unlikely to be found in locations where they would be mistaken for streams. Essentially, the noise can obscure real extensions but it does not create many spurious features. 

\subsubsection{A blind search for simulated streams of unknown parameters}
An even more realistic approach is to vary the properties of the streams so that we do not know if a stream is even present at all. We modify the injected streams to randomly vary their~: (i) direction, so that they are assigned to one of eight different directions (N, NE, E, etc.); (ii) length, ranging from 6 to 8 pixels (we found that below 6 pixels the streams are not recoverable due to the large apparent size of the simulated galaxy); (iii) S/N, ranging from 3.0 to 5.0; (iv) number of channels, from 0 through 5 inclusive so that some streams are not actually injected at all. We also vary the absolute value of the channels used so that the colours of the renzogram provide no information of the presence of a stream. Tests showed that the weakest stream in this parameter space was at the very limit of visual detectability, whereas the strongest is extremely obvious. 

Here we do not impose the strict requirements on any potential detections as in \ref{sec:quicksearch}, but simply try to decide if there is anything present which we might classify as a detection. We allow ourselves to vary the contour level but do not demand that any particular level is necessary - we attempt to select on the same basis as we did for the actual search. This is considerably slower than in \ref{sec:quicksearch} but is far more realistic. As before, we injected 100 potential streams (including those spanning zero channels) into real AGES noise.

Due to the random number of channels selected, a total of 76 streams were actually injected. We recovered 56 of these, giving a completeness of 74\%. We also recorded 5 false positives. All of these were weak compared to the real sources : the strongest being 4\,$\sigma$ in two channels, which is fainter than any of the real streams indentified as ``sure'' or ``probable''. Thus it seems extremely unlikely that any of our ``sure'' and ``probable'' features identified in the real cluster are due to noise : the overwhelming probability is that they are real, physical structures. Again, we emphasise that the reliability level of the ``possible'' detections is surely lower.

\section{Interpreting the streams}
\label{sec:streamint}
\subsection{Comparison with the stripping predictions}
\label{sec:predstrip}
As described in section \ref{sec:koppen}, the advantage of the \cite{koppen} model over \HI{} deficiency is it directly describes current stripping activity. While it is difficult to predefine an exact value of the ratio P$_{loc}$/P$_{req}$ to distinguish active from past strippers, the broader predictions of the model are borne out reasonably well. Of the galaxies with streams of all levels of confidence or noisy contours (11 objects), the median P$_{loc}$/P$_{req}$ ratio is 0.28. Examination of table \ref{tab:allvc1streams} suggests a P$_{loc}$/P$_{req}$ ratio of 0.2 is a plausible (though rough) value to distinguish active from past strippers. Using this in the model, we can state : of the 10 galaxies with detected tails, 8 should be actively losing gas and 2 should be past strippers; of the 9 galaxies without tails, 6 are predicted to be passive and 3 should be active strippers. This broadly supports the assumption that the bulk of the streams should be produced by ram pressure stripping.

While we lack the advantages of resolution from interferometric studies, we can still measure all the same major properties : morphology (at least to say whether a tail is one-sided), length, kinematics, and mass. In our view, all of these tend to support the ram-pressure origin scenario of the tails. They are mostly one-sided, the brightest emission occurs over a relatively short velocity range corresponding to that of the parent galaxy (as in the tails in \citealt{chung}) and they are of comparable length to the \cite{chung} tails, though they are of somewhat lower mass. 

\subsection{Alternative explanations}
\label{sec:predstrip}
\HI{} asymmetries can be produced by internal and external processes. Tidal structures in Virgo (e.g. \citealt{m07}, \citealt{k08}) are usually longer and show more complex, haphazard structures and kinematics, which is also seen in numerical simulations (e.g. \citealt{me17}). Other external processes, such as mergers and accretion, have been invoked in other environments (\citealt{noord}, \citealt{port}) but neither of these is likely in the high velocity dispersion of a cluster.

Could such asymmetrical features be due to internal processes ? Unfortunately comparable studies of isolated galaxies are very rare (three isolated galaxies were targeted with AGES, none of which show extensions - see \citealt{m16}). Even then, the authors can seldom (if ever) rule out external processes (e.g. \citealt{pwl}), and sometimes even find that this is the most likely explanation despite the isolation (\citealt{ramir}). There are few catalogues of isolated galaxies with resolved \HI{} maps, but those which exist broadly suggest that features as long and one-sided as those in the present work are far more likely to result from environmental effects. \cite{noord} note that only 7\% of isolated galaxies are even `mildly' asymmetric. \cite{espa}, using unresolved line profiles, find that the rate of asymmetry is very low in isolation (2\%) compared to the denser field environment (10-20\%). \cite{pwl} describe a VLA survey of 41 isolated galaxies, of which just two (NGC 895 and IC 5078) have clear optically dark \HI{} extensions. The extension of NGC 895 is morphologically very different to the tails described here, while the authors propose that mergers are responsbile for the features in IC 5078. \cite{scott} detect an extension 12 kpc in length, but even for this relatively short feature the authors believe the cause is an external satellite.

In short, while asymmetric extensions are very common in group environments, they are rare in isolation (though there is a lack of of large catalogues to properly quantify this). The only features comparable in both length and morphology to those we have described here are widely attributed to various environmental mechanisms, and we can find no evidence in the literature for internal causes.

\subsection{The evolution of the streams}
\label{sec:streamgobyebye}
Using the \cite{koppen} model, and assuming radial infall, we can estimate the point at which the ram pressure becomes sufficient to start removing gas from the galaxy. We assume this occurs when the ratio P$_{loc}$/P$_{req}$\,$\geq$\,0.5. We can then derive the time of flight from this point to the galaxy's current clustercentric position, allowing us to calculate the dissolution rate. We find rates of typically 1-10\,\Msolar{}\,yr$^{-1}$, with the time since stripping began around 200 Myr. Very similar rates are obtained by the objects described in \citealt{chung}. If these are correct, then the detectable lifetime of the streams is highly variable (simply because of their mass), from a few megayears to a few gigayears. Similar survival times have been calculated independently by \citealt{kent09} and \citealt{boss18b}. The least massive streams are likely in a state of active replenishment, otherwise their lifetime would be so short we would not expect to detect them, whereas the most massive streams can survive long after the stripping event by virtue of their high mass.

\subsection{Relation to the dark clouds}
\label{sec:dudewheresmystream}
\begin{figure*}
\centering
\includegraphics[width=170mm]{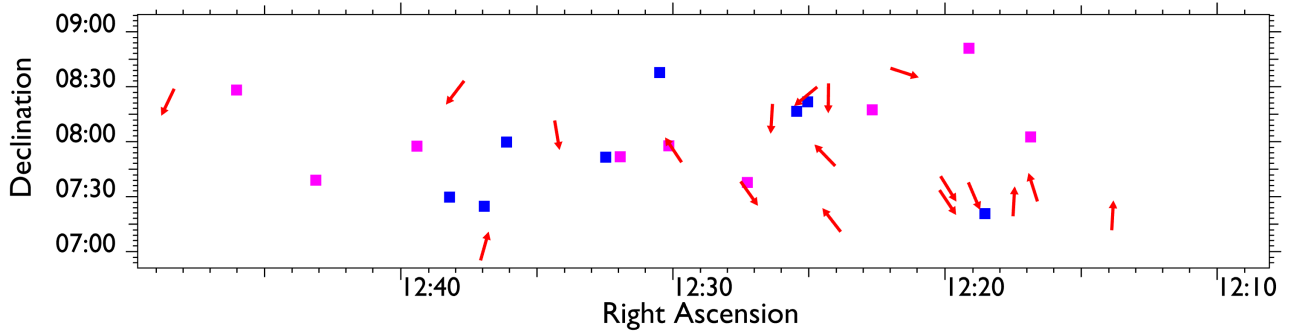}
\caption{Map of all the stream candidates in the VC1 region. Red arrows show the direction the streams point away from their parent galaxies. Blue squares are clouds with no optical counterparts and magenta squares are galaxies with noisy contours but no clear linear extension.}
\label{fig:mainstreammap}
\end{figure*}

Figure \ref{fig:mainstreammap} shows that three of the dark clouds appear to be aligned with two of the streams. In one case (GLADOS 001/AGESVC1 231) the alignment is likely just a projection effect, as the velocity difference between the two is over 700\,\kms{}. In the other case (VCC 740/AGESVC1 247 and 282) the velocity difference between the stream and the clouds is small, however the stream is rather weak. The conclusion of T16 and \cite{me17} that the clouds are unlikely to be the result of gas stripping appears to be sound.

The dissolution rates calculated in section \ref{sec:streamgobyebye} imply that we are witnessing the clouds in the final few Myr of their existence (given their low masses). This is a similar lifetime calculated as in \cite{me17} and \cite{me16}, and considerably shorter than the $\sim$\,100 Myr timescales estimated in \cite{me18}. A caveat is that these rates might strongly vary depending on the density of the objects (i.e. allowing them to self-shield from ionising radiation would reduce evaporation as well as giving them stronger self-gravity), of which we at present do not have direct measurements.

\subsection{Galaxies with noisy contours}
\label{sec:noisy}
\begin{figure}
\centering
\includegraphics[width=85mm]{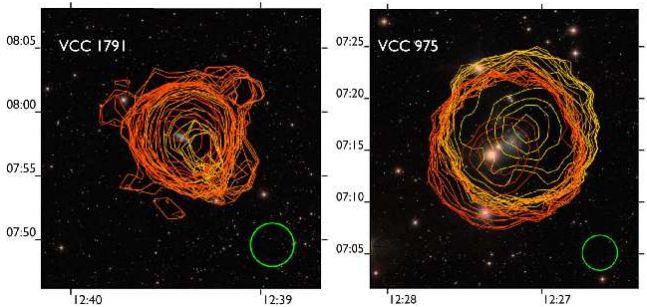}
\caption{Example renzograms of a galaxy identified to have unusually noisy contours (AGESVC1 225, left) compared with another deemed to have clean contours (AGESVC1 277, right). The S/N level of the contours is 5.0 in both cases. Both are bright sources with peak S/N levels of 113 for AGESVC1 225 and 130 for AGESVC1 277. Note that since both sources are marginally resolved, a slight position shift in the centre of their contours can be seen, especially for VCC 975.}
\label{fig:noisymcnoiseface}
\end{figure}

Figure \ref{fig:noisymcnoiseface} provides a comparison of galaxies with smooth and irregular contours. It is possible that in the weaker cases some of these irregular features are not real, but some cases, which are seen at 5$\sigma$ in many pixels and channels, are unambiguous. These galaxies show no obvious coherent distribution within the cluster (figure \ref{fig:mainstreammap}). One possibility is that they are actively losing gas but with motion mainly along the line of sight. Another is that they are losing gas by different, internal mechanisms, e.g. that the extensions are only due to temporary gas displacement (not complete ejection) via supernovae, for example; many dwarf galaxies show irregular contours (\citealt{swat}). While mergers are not expected in the cluster environment, IC 5078 in \cite{pwl} (see their figure 40) shows intriguingly similar contours to these `noisy' galaxies, which the authors attribute to a possible minor merger.

\section{Summary and Discussion}
\label{sec:conc}
While approximately 60\% of the late-type galaxies in Virgo show significant \HI{} deficiencies, only 2\% have previously documented streams. We might expect to see more galaxies in the process of actively losing gas, but quantification is difficult as many different factors influence the exact number of expected streams~: how many galaxies are currently stripping, the detectability of the streams in \HI{} surveys given the stream geometry, orientation and kinematics, and the gas phase change rate.

We found through a geometric model that the orientation of the streams is unlikely to significantly reduce the detected number, assuming that previous detections are representative of the full population. Based on existing detections and the geometrical correction, we estimated that ALFALFA should detect 46 streams and AGES 11. Additionally, the model of \cite{koppen} predicts that streams should tend to be associated with galaxies of higher P$_{loc}$/P$_{req}$ (the ratio of predicted local ram pressure and required pressure to explain the observed \HI{} deficiency), and while not every galaxy with a high pressure ratio should have a tail, most galaxies with tails should have high ratios.

We re-examined the two AGES Virgo cubes using upgraded data processing and visualisation techniques. We found a minimum of 10 streams (only 1 of which was previously known) and potentially as many as 26. This included galaxies observed in VIVA for which only much shorter tails were previously seen, suggesting that our higher detection rate is in part due to the greater sensitivity of AGES. We demonstrated using statistical analyses that our 10 most confident detections are too bright, extended and coherent to be a result of noise. The predictions from the \cite{koppen} model appear to hold true. Furthermore we can make a quantitative prediction for future surveys that streams will be most commonly associated with galaxies with P$_{loc}$/P$_{req}$\,$\gtrsim$\,0.2. 

Several factors have contributed to the puzzling lack of previous stream detections, most notably quantitative estimates of~: 1) the number of currently active stripping galaxies; 2) the time galaxies have spent thus far in the cluster; 3) the evaporation and dispersal rate of stripped material; and additionally 4) the discovery of previously unknown streams. While a naive comparison of the number of deficient galaxies and those with streams shows a clear mismatch, the more detailed analysis reveals a much better agreement.

While a few Virgo cluster streams contain a significant fraction of the missing gas, most contain only a few percent of the missing gas mass of their parent galaxy (in contrast, streams in other environments often contain an appreciable fraction of the \HI{} mass of the parent, as discussed in \citealt{me14}). This strongly suggests that in most cases a phase change occurs during the stripping process that renders much of the stream undetectable to \HI{} surveys. There is supporting evidence for this in other clusters : \cite{gv18} find that 50-60\% of observed late-type galaxies in the Coma cluster have H$\alpha$ tails.

The \cite{koppen} model allows a prediction of the time each galaxy has been undergoing active stripping. For the galaxies with streams this allows us to estimate the dissolution time independently of the physical processes at work, which are typically 1-10\,\Msolar{}/yr. In combination with the time spent within the cluster thus far, these rapid dissolution rates are consistent with the existence of both a few long, relatively massive streams, and a larger population of shorter, less massive tails. Massive streams can persist for longer due to their greater mass, whereas less massive streams are only detectable due to constant replenishment as material is still being stripped from their parent galaxy. The dissolution rates also suggest that, if the optically dark isolated clouds are of a similar nature to the streams, we must be witnessing them in a very short detection window and they should disappear in the next few Myr.

We expect to detect a similar number of streams with WAVES, which has a similar sensitivity and area of coverage to the AGES Virgo fields (\citealt{m19}). Better statistics will allow further tests of the stream models and place more constraints on the nature of the optically dark clouds, but our information about the features described here is limited by the resolution of Arecibo. Future observations at comparable column density sensitivty and higher resolution may give much more precise information on the nature and formation mechanism of individual streams.

\section*{Acknowledgments}
We are grateful to the anonymous referee whose comments improved the manuscript.

This work was supported by the Czech Ministry of Education, Youth and Sports from the large lnfrastructures for Research, Experimental Development and Innovations project LM 2015067, the Czech Science Foundation grant CSF 19-18647S, and the institutional project RVO 67985815.

This work is based on observations collected at Arecibo Observatory. The Arecibo Observatory is operated by SRI International under a cooperative agreement with the National Science Foundation (AST-1100968), and in alliance with Ana G. M\'{e}ndez-Universidad Metropolitana, and the Universities Space Research Association.

The SOFIA Science Center is operated by the Universities Space Research Association under NASA contract NNA17BF53C.

This research has made use of the GOLDMine Database. This research has made use of the NASA/IPAC Extragalactic Database (NED) which is operated by the Jet Propulsion Laboratory, California Institute of Technology, under contract with the National Aeronautics and Space Administration.

This work has made use of the SDSS. Funding for the SDSS and SDSS-II has been provided by the Alfred P. Sloan Foundation, the Participating Institutions, the National Science Foundation, the U.S. Department of Energy, the National Aeronautics and Space Administration, the Japanese Monbukagakusho, the Max Planck Society, and the Higher Education Funding Council for England. The SDSS Web Site is http://www.sdss.org/.

The SDSS is managed by the Astrophysical Research Consortium for the Participating Institutions. The Participating Institutions are the American Museum of Natural History, Astrophysical Institute Potsdam, University of Basel, University of Cambridge, Case Western Reserve University, University of Chicago, Drexel University, Fermilab, the Institute for Advanced Study, the Japan Participation Group, Johns Hopkins University, the Joint Institute for Nuclear Astrophysics, the Kavli Institute for Particle Astrophysics and Cosmology, the Korean Scientist Group, the Chinese Academy of Sciences (LAMOST), Los Alamos National Laboratory, the Max-Planck-Institute for Astronomy (MPIA), the Max-Planck-Institute for Astrophysics (MPA), New Mexico State University, Ohio State University, University of Pittsburgh, University of Portsmouth, Princeton University, the United States Naval Observatory, and the University of Washington.

{}

\clearpage

\appendix
\counterwithin{figure}{section}
\section{Comments on individual streams}
\label{sec:appendix}
In this section we describe the individual streams in more detail. We also provide figures of each stream not given in the main text (for those not shown here, see figures \ref{fig:goodstreams}, \ref{fig:vcc1205} and \ref{fig:noisymcnoiseface}). All figures are renzograms unless otherwise stated, with nearby galaxies labelled and the target galaxy unlabelled in the centre of the figure. We also give the contour level shown in the renzogram (for reference, 1$\sigma$ is equivalent to $N_{HI}$\,=\,1.5$\times$10$^{17}$\,cm$^{-2}$) for each galaxy figure and the exact velocity range used for each renzogram.

\subsubsection{VCC 2070}
\begin{figure*}
\begin{center}  
  \subfloat[VCC 2070]{\includegraphics[width=85mm]{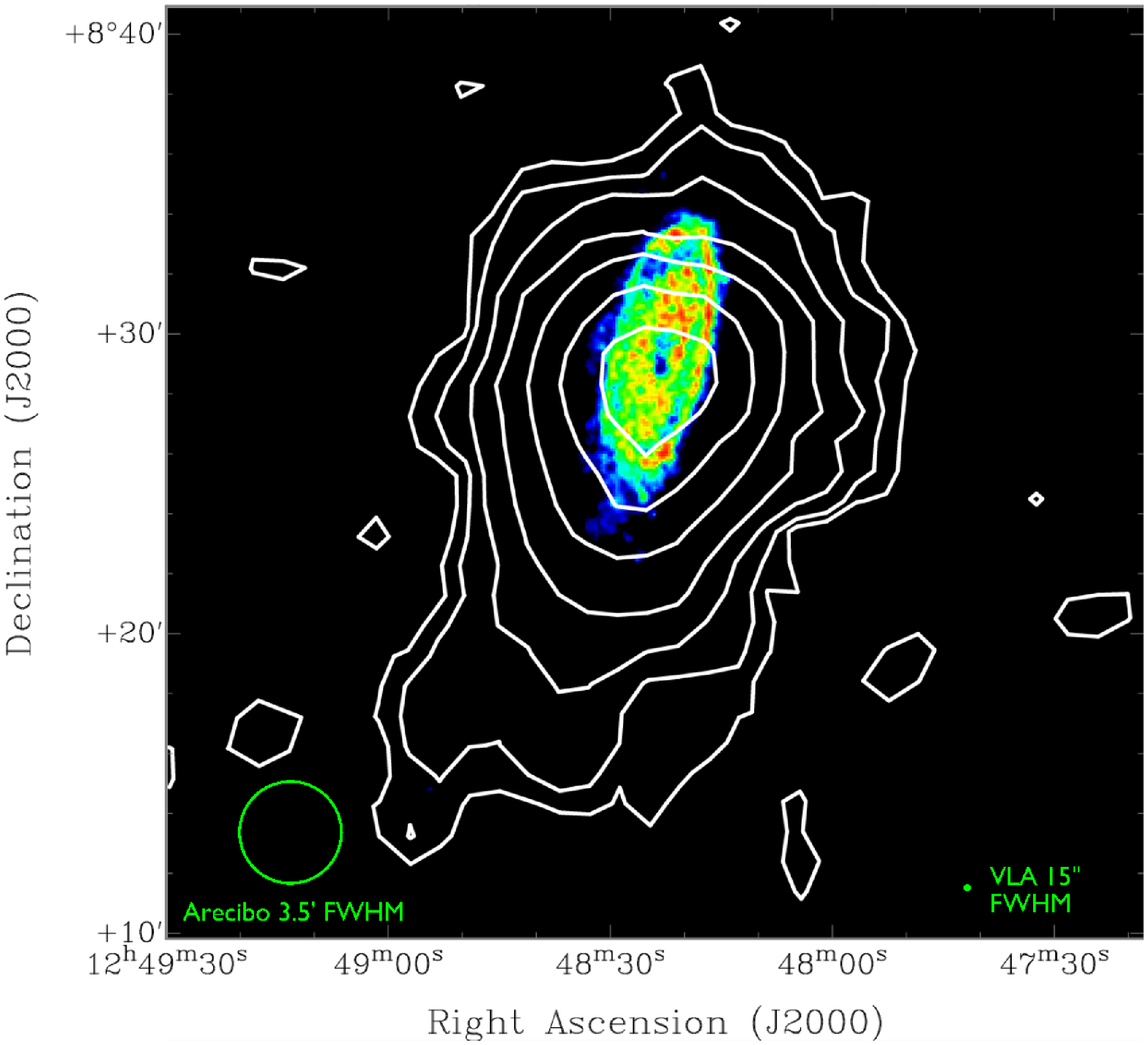}}
  \subfloat[VCC 1555]{\includegraphics[width=85mm]{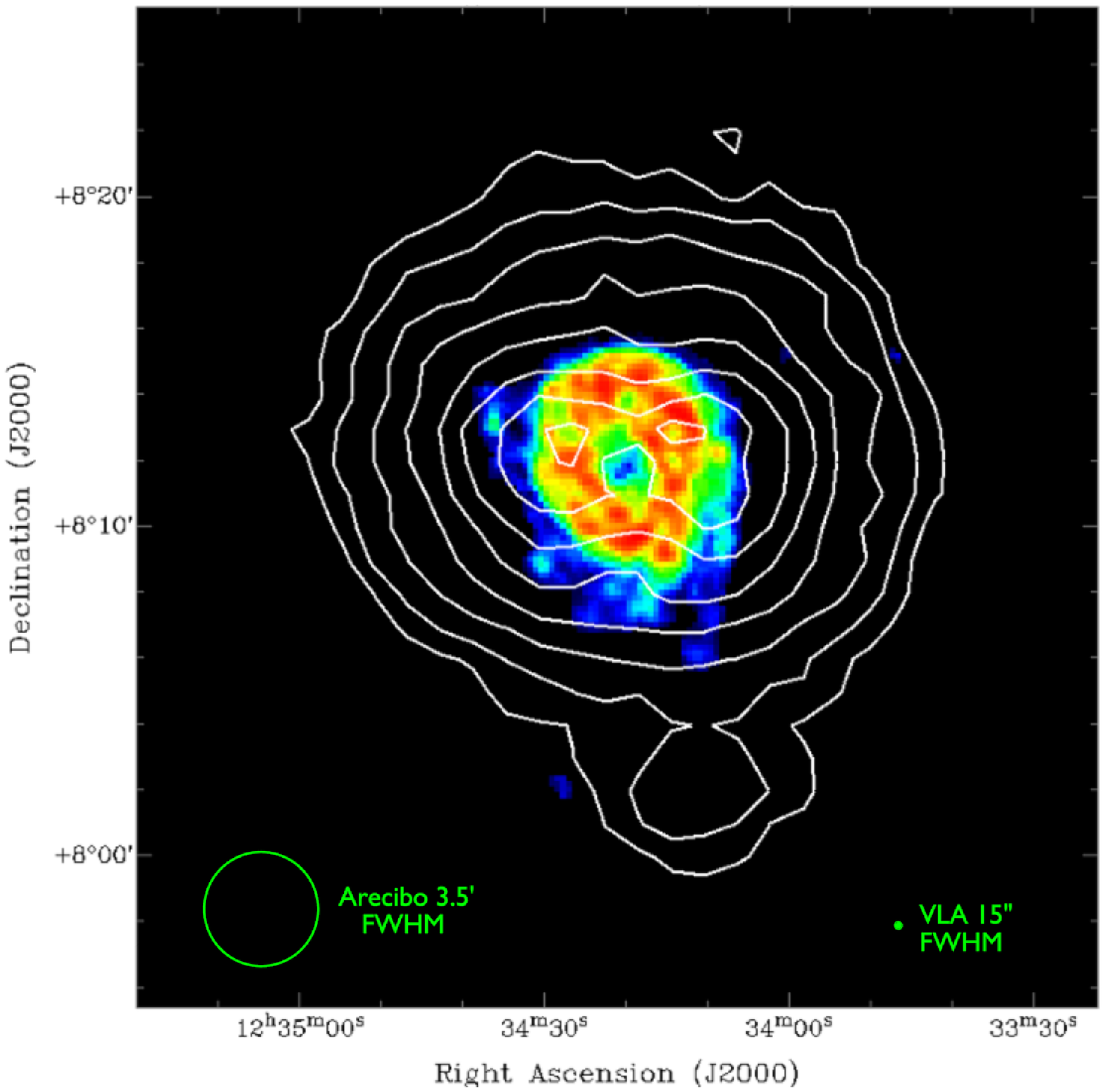}} 
\caption{The two galaxies detected in the AGES VC1 data cube which were also mapped with the VIVA survey. The RGB colours show the VIVA moment 0 maps while the (logarithmically spaced) contours show the moment 0 maps from AGES.}
\label{fig:vivamaps}
\end{center}
\end{figure*}

This is our clearest detection of a stream, and perhaps the most surprising. The galaxy is near the edge of cube and about 1.7 Mpc in projection from M87. The edge of the stream (at 4$\sigma$) reaches 4 Arecibo beams from the centre of the galaxy, giving it a maximum extent of about 70 kpc. The stream is visible in at least 12 velocity channels and peaks at around 8$\sigma$ above the noise. This galaxy is one of only two in the VC1 area observed with VIVA. Although the long stream is not detected, the VIVA moment 0 map clearly shows a ragged edge (with a hint of a short extension) on the same side of the galaxy as the long AGES stream, with the opposite edge being much smoother - see figure \ref{fig:vivamaps}.

This detection clearly demonstrates the capacity of the AGES observations to detect features not visible to VIVA because of its comparatively poor column density sensitivity, bearing in mind the caveats discussed in section \ref{sec:geometry}. The main reason this stream appears to have been missed in our original search is probably beam smearing - as shown in figure \ref{fig:renzoexample}, the stream appears obvious \textit{only} if the data is processed in the correct way, with the use of contour maps greatly enhancing the non-circular \HI{} extension.

This galaxy is shown in figure \ref{fig:goodstreams} with the contour at 4$\sigma$ over the velocity range 1,102 - 1,242\,\kms{}.

\subsubsection{VCC 1555}
This is our second unambiguous detection of a stream. It appears in figure \ref{fig:goodstreams} (and in the data cube) almost to resemble a companion galaxy, but we can find no optical counterpart at the position of the extended contours. We searched both the SDSS data (smoothing the FITS files so as to increase sensivity to low surface brightness features) and the deeper NGVS data (\citealt{ngvs}).

Since this is an exceptionally bright galaxy with a peak S/N of 140, beam smearing is extremely strong and makes it difficult to estimate the length and velocity width of the stream. The furthest point of the extension is about 3.5 Arecibo beams at 4$\sigma$, or 60 kpc, from the centre of VCC 1555. It appears to be present in 25 velocity channels, though we stress this is uncertain as it is difficult to rigorously define where the disc ends and the stream begins (in some channels it is unclear if the `stream' is actually entirely detached from the disc). The peak flux of the stream is approximately 14$\sigma$. This galaxy was also observed with VIVA, and as with VCC 2070 there are hints of a short extension visible in the VIVA data roughly corresponding to the position of the long AGES stream, though the alignment is not quite as good as the case of VCC 2070 (see figure \ref{fig:vivamaps}).

This galaxy is shown in figure \ref{fig:goodstreams} with the contour at 5$\sigma$ over the velocity range 1948 - 2021\,\kms{}.

\subsubsection{VCC 2062/2066}
\begin{figure}
\centering
\includegraphics[width=85mm]{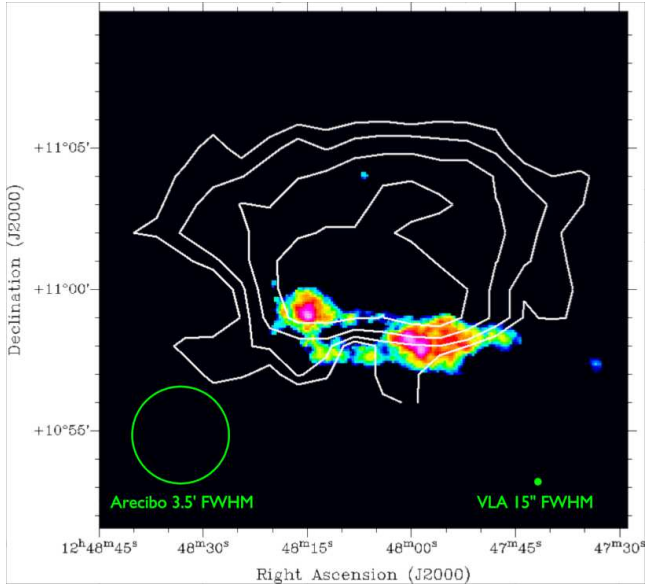}
\caption{VCC 2062/2066, the single system in VC2 detected by both AGES and VIVA, shown as in figure \ref{fig:vivamaps}. A distortion can be seen in the contours south of declination 10:59:00, where the noise level of the data increases significantly near the edge of the cube.}
\label{fig:vcc2066}
\end{figure}

While the stream connecting VCC 2062/2066 is well known and already seen in AGES data (\citealt{me13}), accurate measurements are difficult since the feature is very close to the southern limit of the data. It appears to be significantly more extended than in VIVA data but not more elongated, as shown in figure \ref{fig:vcc2066}. While the \cite{koppen} model predicts the galaxy should be stripping, the origin of the gas in this case is less clear. VCC 2062 is too small, VCC 2066 is a lenticular (which are generally gas-poor in Virgo, see \citealt{me12} and \citealt{me13}) - and the bulk of the gas is offset from VCC's stellar component. A detailed discussion on this system is given in \citealt{duc07}.

This galaxy is shown in figure \ref{fig:goodstreams} with the contour at 4$\sigma$ over the velocity range 1,088 - 1,291\,\kms{}.

\subsubsection{VCC 888}
This is a borderline sure/possible case as the stream is relatively small and only detected in 3-4 channels, with a peak flux of only 5$\sigma$, though the extension itself appears clearly different from the otherwise clean contours. The stream extends to about 2 Arecibo beams from the optical centre of the galaxy, or 45 kpc at the GOLDMine distance of 23 Mpc.
% In figure 7

This galaxy is shown in figure \ref{fig:goodstreams} with the contour at 4$\sigma$ over the velocity range 1,087 - 1,123\,\kms{}. 

\subsubsection{VCC 1699}
This galaxy is at the very southern edge of the AGES cube so unfortunately we cannot examine the whole galaxy. The extension is short, no more than 2 beams (35 kpc at 17 Mpc distance) but detected at the 7-8$\sigma$ level. It appears to be somewhat less linear than the other features : while the peak S/N levels are found in the extension leading to the north-west, a possible north-eastern extension is detected at the 6$\sigma$ level. The north-western extension is detected in at least 8 velocity channels.
% In figure 7

This galaxy is shown in figure \ref{fig:goodstreams} with the contour at 5$\sigma$ over the velocity range 1,556 - 1,735\,\kms{}. 

\subsubsection{VCC 393}
The extension here appears to be quite linear. It is relatively weak with a peak detection of 5$\sigma$, found in 4 velocity channels. Its maximum extent is around 2 Arecibo beams or 45 kpc at its assumed 23 Mpc distance. The galaxy appears to be somewhat optically disturbed.
% In figure 7

This galaxy is shown in figure \ref{fig:goodstreams} with the contour at 4$\sigma$ over the velocity range 2,513 - 2,723\,\kms{}.

\subsubsection{VCC 94}
While the contours of this galaxy are roughly circular for most of its velocity range, over about 10 consecutive velocity channels they are distinctly ellipsoidal, with a coherent north-south alignment. This structure is detected at a peak flux equivalent to 25$\sigma$ in some channels; only the effects of beam smearing (as the galaxy itself is a bright source) caution us to give it a `sure' rather than `certain' detection rating. At the 4$\sigma$ level, this source is extended up to 2.5 Arecibo beams from the galaxy's optical centre, equivalent to 80 kpc at its 32 Mpc distance.
% In figure 7

This galaxy is shown in figure \ref{fig:goodstreams} with the contour at 6$\sigma$ over the velocity range 2,493 - 2,587\,\kms{}.

\subsubsection{EVCC 2234 (AGESVC2 025 A and B) and NGC 4746}
\begin{figure}
\centering
\includegraphics[width=85mm]{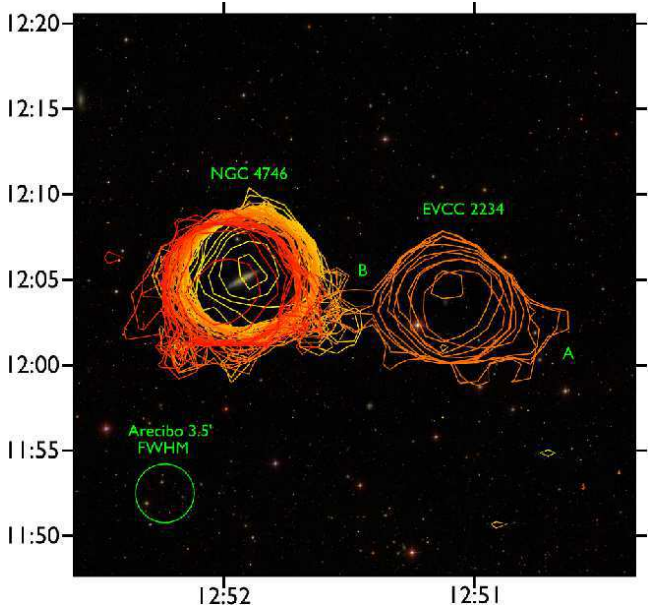}
\caption{Feature 'B' in the image is a possible bridge between the bright spiral NGC 4746 and the fainter irregular EVCC 234.}
\label{fig:evcc284}
\end{figure}

The western extension of EVCC 2234 is not very large, but the non-circular contours persist over 3 or 4 channels and are clearly seen at 5$\sigma$. We regard this as a reasonly secure detection This extension was already shown in figure \ref{fig:goodstreams} but figure \ref{fig:evcc284} reveals two other, more tenatative detections in the same system. EVCC 2234 appears to be interacting with the bright spiral NGC 4746, which is just outside the VCC catalogue area but almost certainly a cluster member itself. There is a hint of a possible bridge between the two galaxies, but it is present over only a few channels so we regard this as rather tenative. The south-east extension from NGC 4746 appears as asymmetrical noisy contours over $\sim$\,20 channels so is more secure. Given the identical line of sight velocities of the two galaxies, they are likely interacting.

Both of these galaxies are shown in figure \ref{fig:goodstreams} with the contour at 4$\sigma$ over the velocity range 1,574 - 1,991\,\kms{}.

\subsubsection{VCC 1859}
In contrast to the EVCC 2234 / NGC 4746 pair, the VCC 1859/1868 pair do not appear to be interacting. The velocity difference of the two is 630 \kms{}, high but not necessarily forbidding an interaction given the velocity dispersion of the cluster. VCC 1868 shows clean, circular contours, whereas VCC 1859 (a 12$\sigma$ detection) shows elongated, non-circular contours. The extension is a somewhat tenatative detection given its small size.

This galaxy is shown in figure \ref{fig:goodstreams} with the contour at 4$\sigma$ over the velocity range 1,596 - 1,746\,\kms{}.

\subsubsection{VCC 1205}
\label{sec:VCC1205}
\begin{figure}
\centering
\includegraphics[width=85mm]{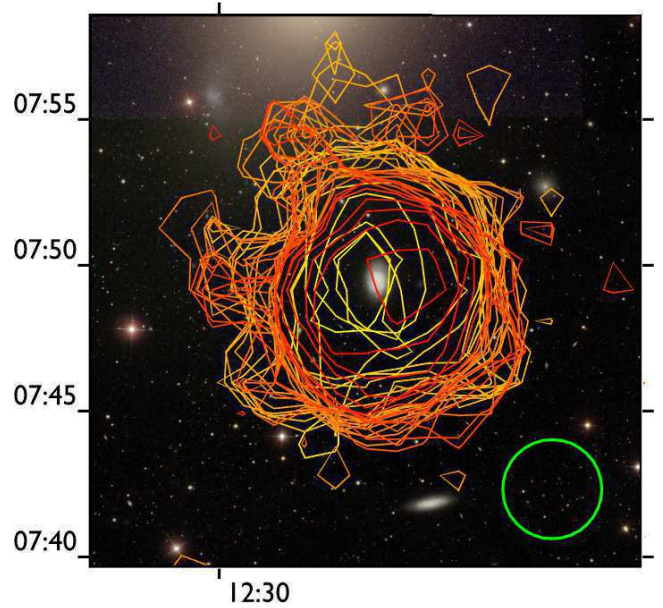}
\caption{Renzogram of AGESVC1 212 (VCC 1205), which seems to have more irregular contours on the north-western side compared to the south east. Contours are at the 4$\sigma$ level.}
\label{fig:vcc1205}
\end{figure}

This is a borderline noisy/sure case - the galaxy certainly has noisy contours, easily visible at 4 sigma in figure \ref{fig:vcc1205}, but they are distinctly one-sided. We have therefore assumed in the text that this galaxy is actually another case of a tail, albeit a rather ragged one compared to the others. Contours in the figure are shown over the velocity range 2,184 - 2,475\,\kms{}.

\subsubsection{VCC 1249}
\label{sec:VCC1249}
This is a unique case of displaced gas, discussed in T12 section 4.7 and figure 21. Briefly, while no gas is detected at the optical position of VCC 1249 itself (table \ref{tab:virgoggalas} gives the upper limit from AGES), an \HI{} cloud (AGESVC1 281, though it was previously detected in many other observations) is found midway between VCC 1249 and the nearby M49. Tables \ref{tab:virgogas} and \ref{tab:allvc1streams} give the \HI{} mass for the cloud. The expected \HI{} mass for the galaxy is 4.5$\times$10$^{8}$\Msolar{}, so the cloud accounts for only about 10\% of the missing gas.

\subsubsection{VCC 667}
\begin{figure}
\centering
\includegraphics[width=85mm]{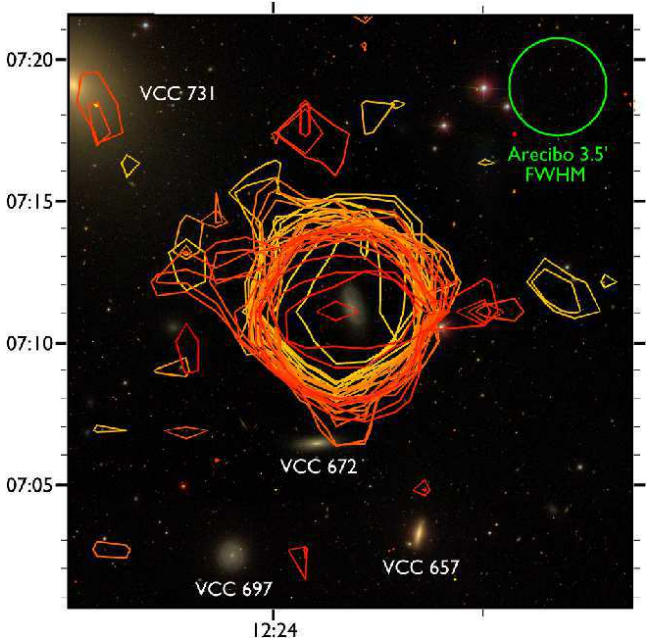}
\caption{VCC 667 (centre) and immediate environs.}
\label{fig:vcc667}
\end{figure}

This is a weak stream only visible at 3$\sigma$ (figure \ref{fig:vcc667}). Several other galaxies are found in its immediate vicinity : VCC 657 (760 \kms{}), 672 (922 \kms{}), 697 (1,230 \kms{}) and 731 (1,242 \kms{}). The positional alignment of VCC with the stream and its close velocity match to VCC 667 (1,405 \kms{}) suggest a possible tidal interaction .

This galaxy is shown in figure \ref{fig:vcc667} with the contour at 3$\sigma$ over the velocity range 1,313 - 1,542\,\kms{}.

\subsubsection{AGESVC1 204}
\begin{figure}
\centering
\includegraphics[width=85mm]{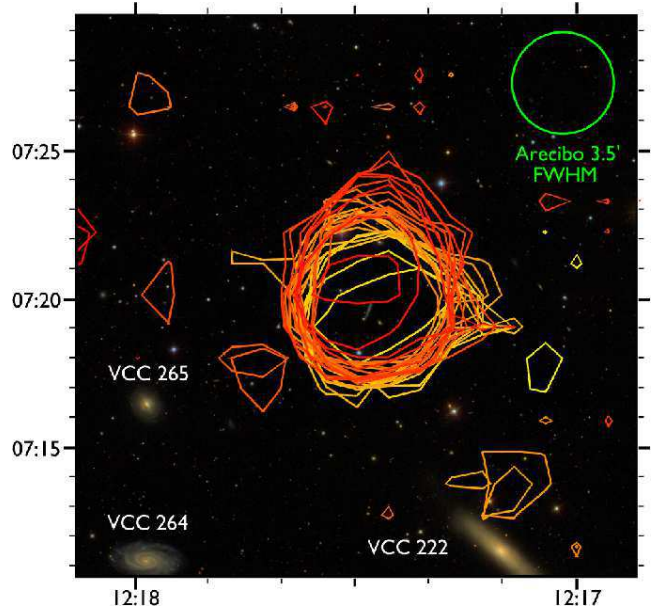}
\caption{AGESVC1 204 (centre) and immediate environs.}
\label{fig:agesvc1204}
\end{figure}

A flat, edge-on blue galaxy with non-circular contours (visible at 5$\sigma$) that are elongated north-south (figure \ref{fig:agesvc1204}). Two of the nearby galaxies (VCC 265 and 264) have velocities $>$\,3,000 \kms{} and are therefore not physically associated with the galaxy. VCC 222 is at 2,410 \kms{} whereas AGESVC1 204 is at 2,210 \kms{}, suggesting a possible physical association though the tail from VCC 204 is rather short and one-sided.

This galaxy is shown in figure \ref{fig:agesvc1204} with the contour at 3$\sigma$ over the velocity range 2,121 - 2,299\,\kms{}.

\subsubsection{VCC 199}
\begin{figure}
\centering
\includegraphics[width=85mm]{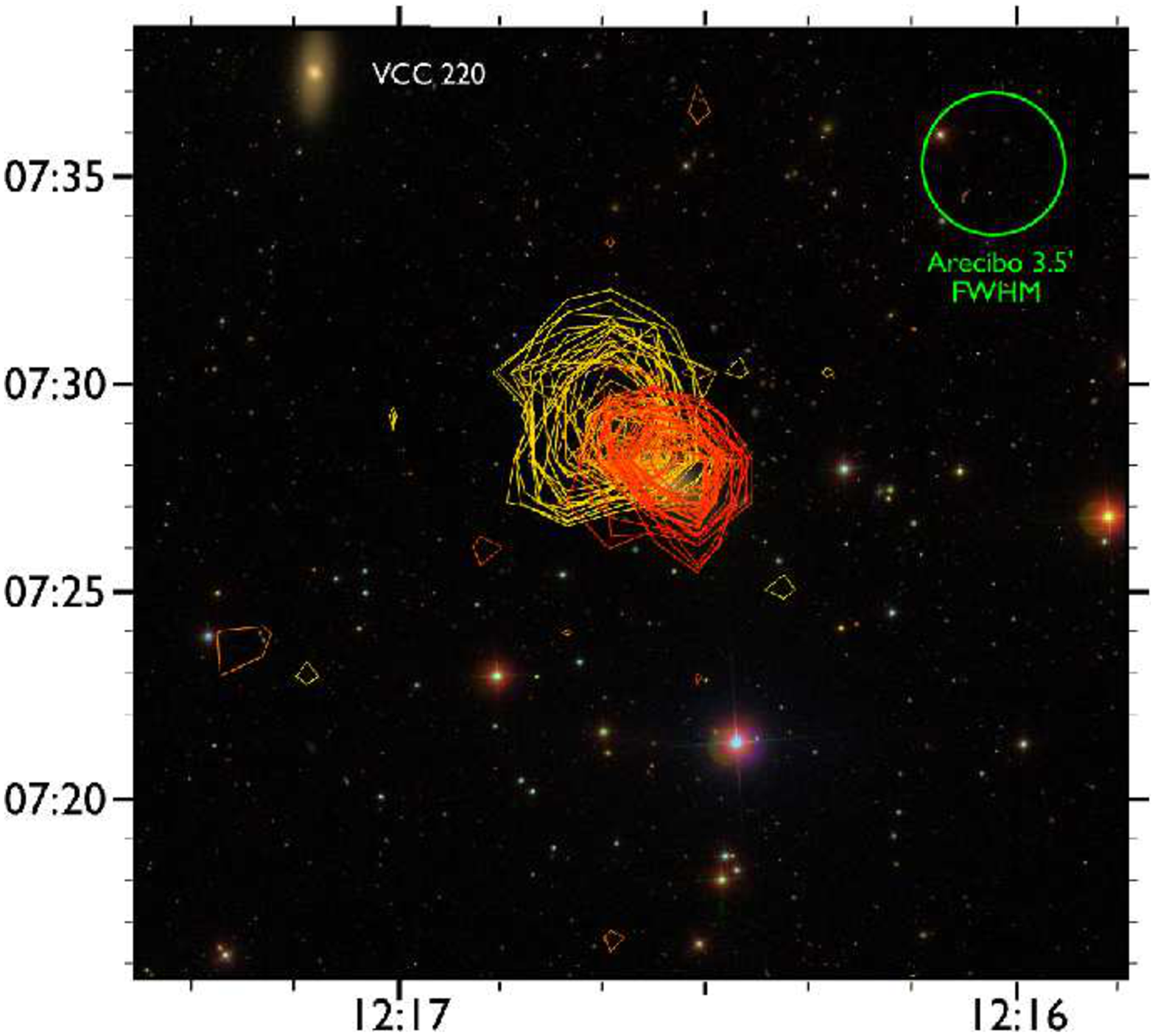}
\caption{VCC 199.}
\label{fig:VCC199}
\end{figure}

VCC 199 is an early-type spiral with a relatively weak \HI{} detection (figure \ref{fig:VCC199}). At 4$\sigma$, its contours tentatively suggest a short extension to the north-east, however, it may be that the galaxy is simply marginally resolved with a slightly asymmetric velocity profile. However the nearby galaxy VCC 222 is at a similar velocity (2,306 \kms{}) to VCC 199 (2,603 km/s) and well-aligned with the long axis of the possible extension.

This galaxy is shown in figure \ref{fig:VCC199} with the contour at 5$\sigma$ over the velocity range 2,340 - 2,896\,\kms{}.

\subsubsection{AGESVC1 232}
\begin{figure}
\centering
\includegraphics[width=85mm]{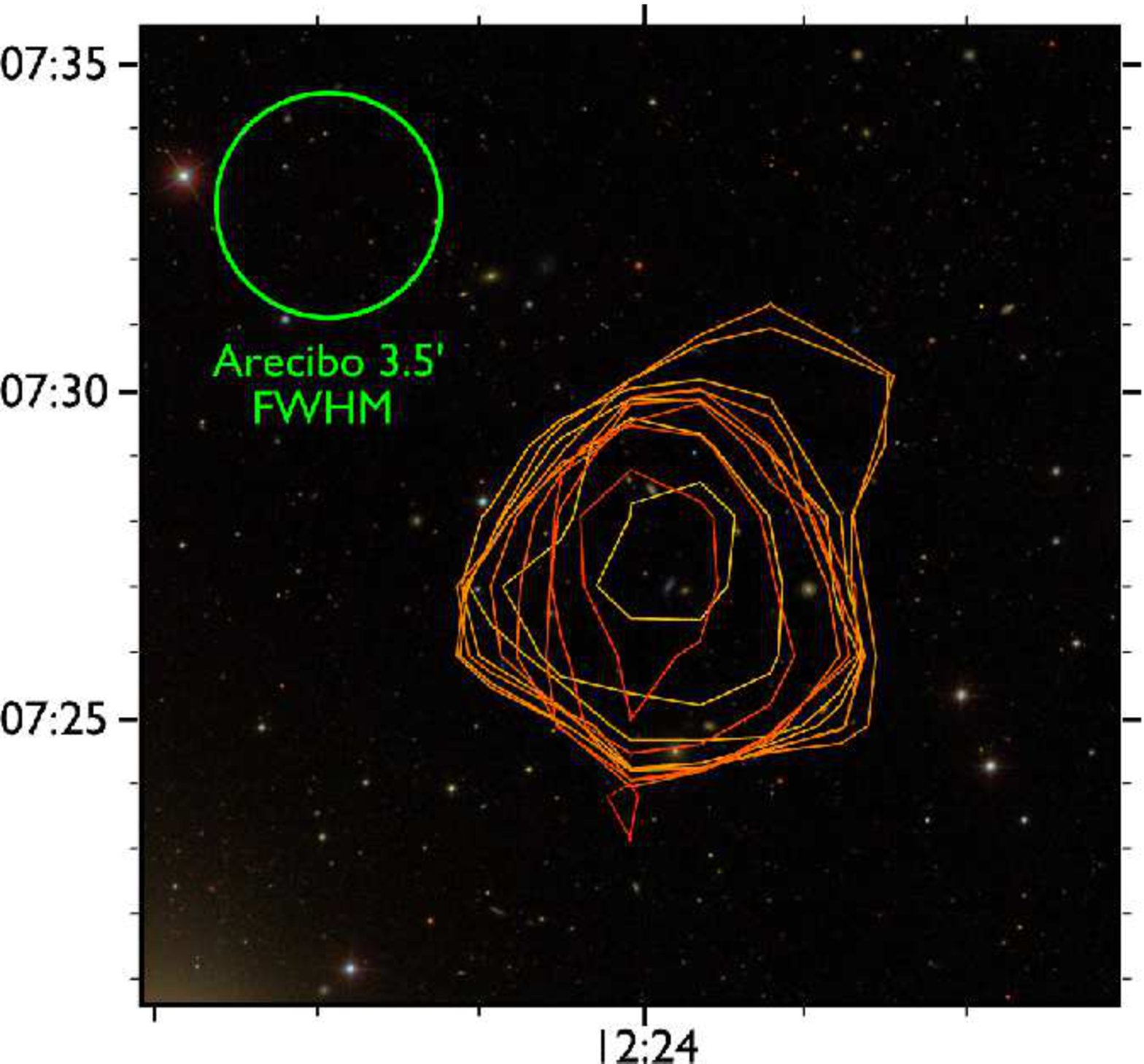}
\caption{AGESVC1 232}
\label{fig:AGESVC1232}
\end{figure}

This is a small blue dwarf galaxy with a possible north-west extension (figure \ref{fig:AGESVC1232}). This is a very uncertain detection as it is short, weak (4$\sigma$) and only present in a couple of velocity channels.

This galaxy is shown in figure \ref{fig:AGESVC1232} with the contour at 4$\sigma$ over the velocity range 1,196 - 1,295\,\kms{}.

\subsubsection{VCC 1011}
\begin{figure}
\centering
\includegraphics[width=85mm]{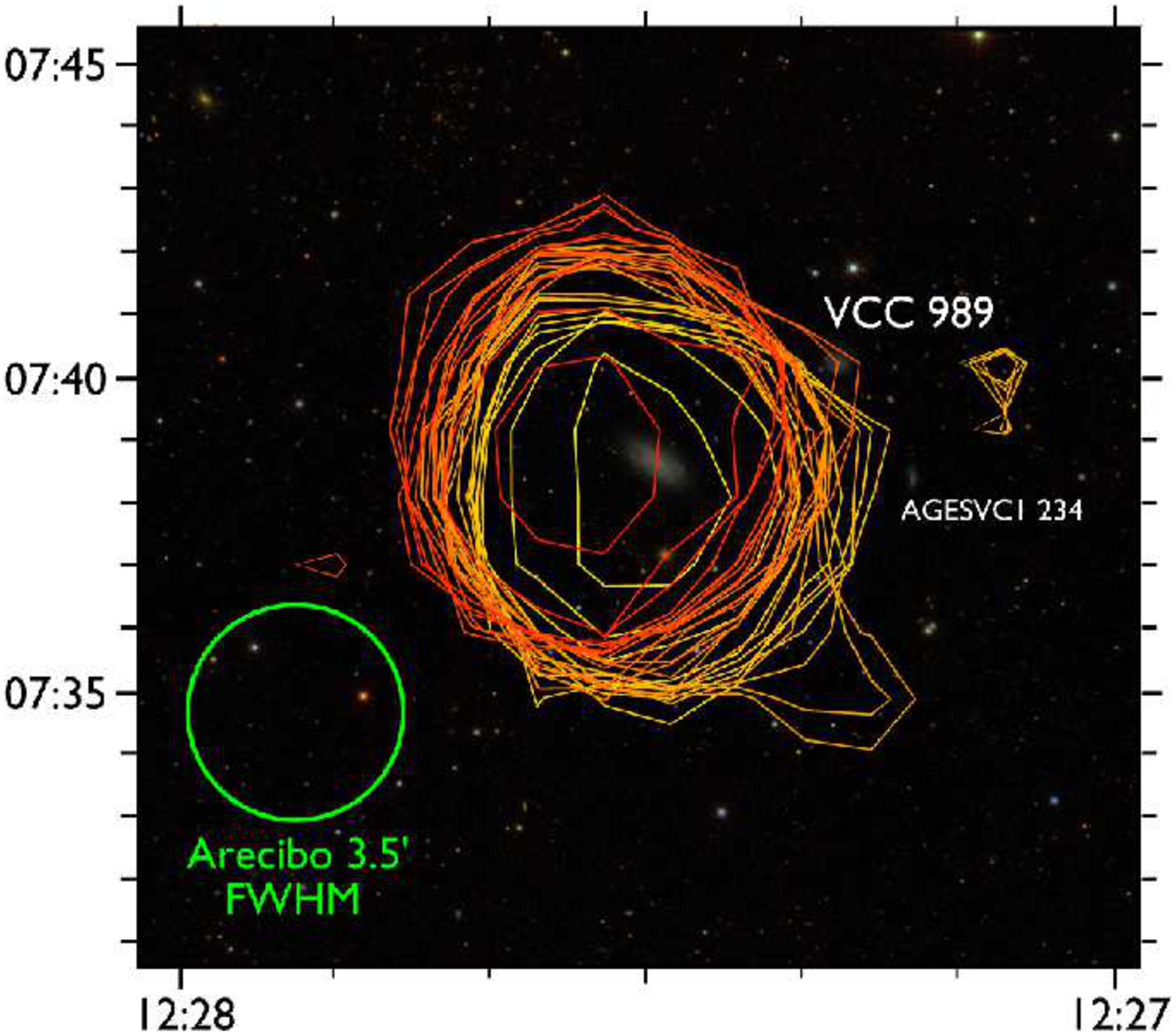}
\caption{VCC 1011}
\label{fig:VCC1011}
\end{figure}

VCC 1011 is a spiral galaxy with a possible extension aligned with the plane of its disc (figure \ref{fig:VCC1011}). The extension is rather pronounced, but only visible at 4$\sigma$ in a few channels. The nearby galaxy VCC 989 is at 1,846 \kms{} so unlikely to be associated with VCC 1011 (at 867 \kms{}). The dwarf galaxy AGESVC1 234 is also nearby and rather closer in velocity at 1,184 \kms{} but shows no sign of any extensions itself.

This galaxy is shown in figure \ref{fig:VCC1011} with the contour at 4$\sigma$ over the velocity range 786 - 968\,\kms{}.

\subsubsection{VCC 688}
\begin{figure}
\centering
\includegraphics[width=85mm]{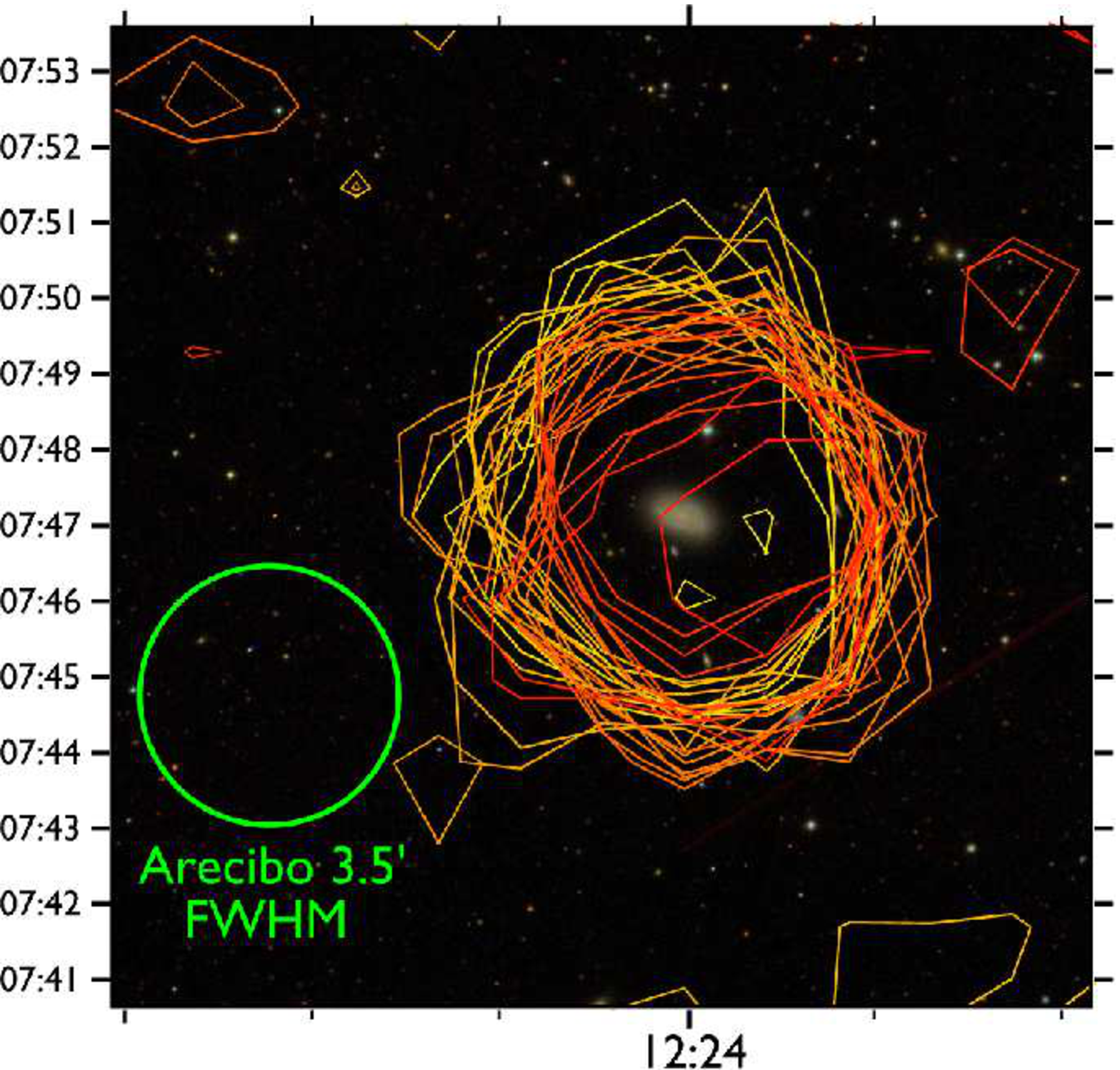}
\caption{VCC 688}
\label{fig:VCC688}
\end{figure}

This spiral galaxy shows a 3-4$\sigma$ hint of asymmetrically noisy contours on its north-eastern side (figure \ref{fig:VCC688}). It is 20$^{\prime}$ (100 kpc) due north of AGESVC1 232, which as mentioned also has a hint of an extension and is at a similar velocity, but no other galaxies are visible nearby.

This galaxy is shown in figure \ref{fig:VCC688} with the contour at 3$\sigma$ over the velocity range 1,043 - 1,230\,\kms{}.

\subsubsection{VCC 740}
\begin{figure}
\centering
\includegraphics[width=85mm]{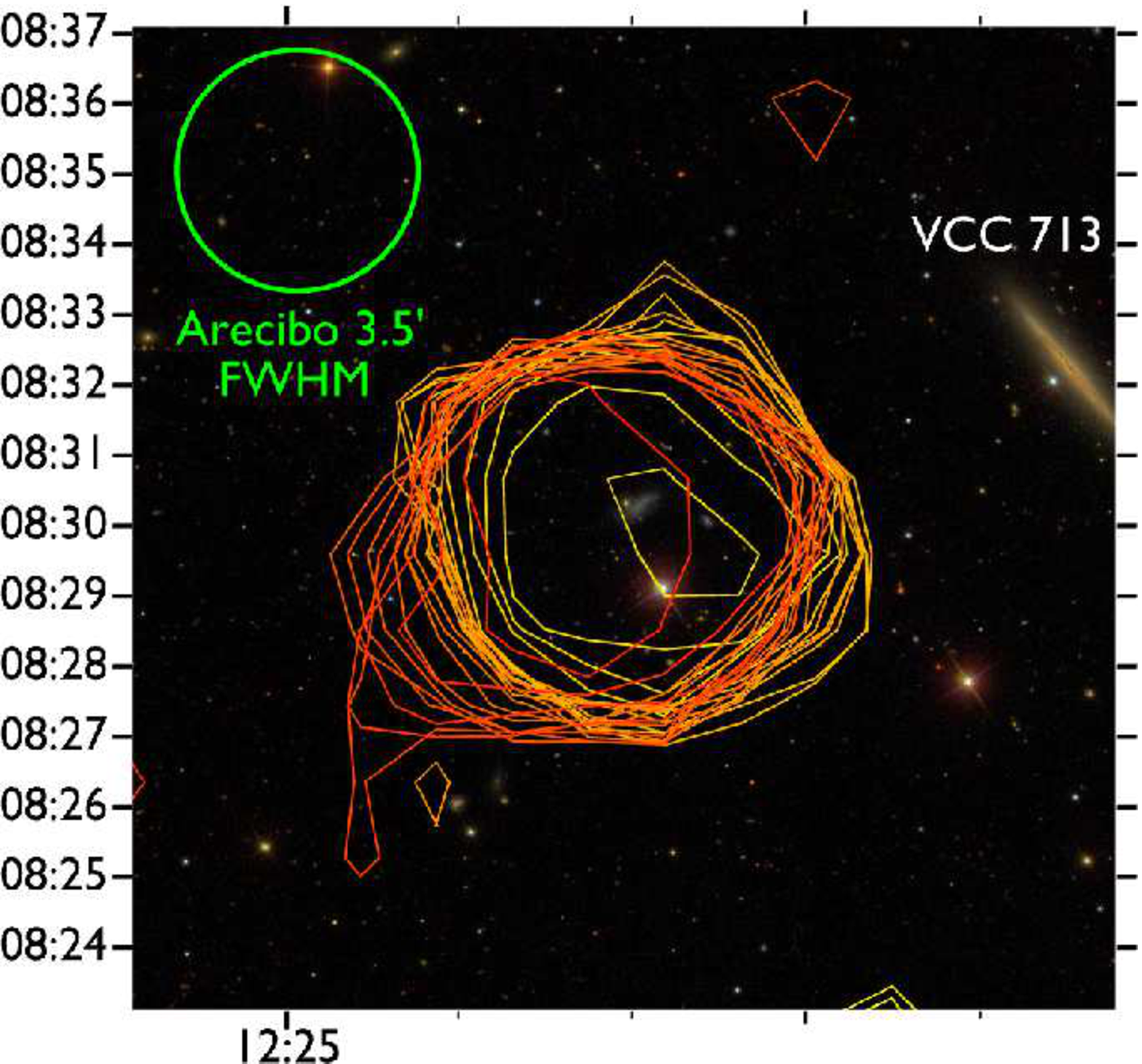}
\caption{VCC 740}
\label{fig:VCC740}
\end{figure}

The extension on this irregular galaxy (figure \ref{fig:VCC740}) is visible at 4$\sigma$ with hints at 5$\sigma$. It appears to be significantly large and asymmetrical in comparison to the galaxy itself but is only visible over a few velocity channels. The bright spiral VCC 713, which has its own hint of an extension, is visible nearby. VCC 713 is at 1,138 \kms{} whereas VCC 740 is at 877 \kms{}. The stream of VCC 713 runs due south, making it unclear if these two galaxies are interacting.

This galaxy is shown in figure \ref{fig:VCC740} with the contour at 4$\sigma$ over the velocity range 797 - 968\,\kms{}.

\subsubsection{VCC 713}
\begin{figure}
\centering
\includegraphics[width=85mm]{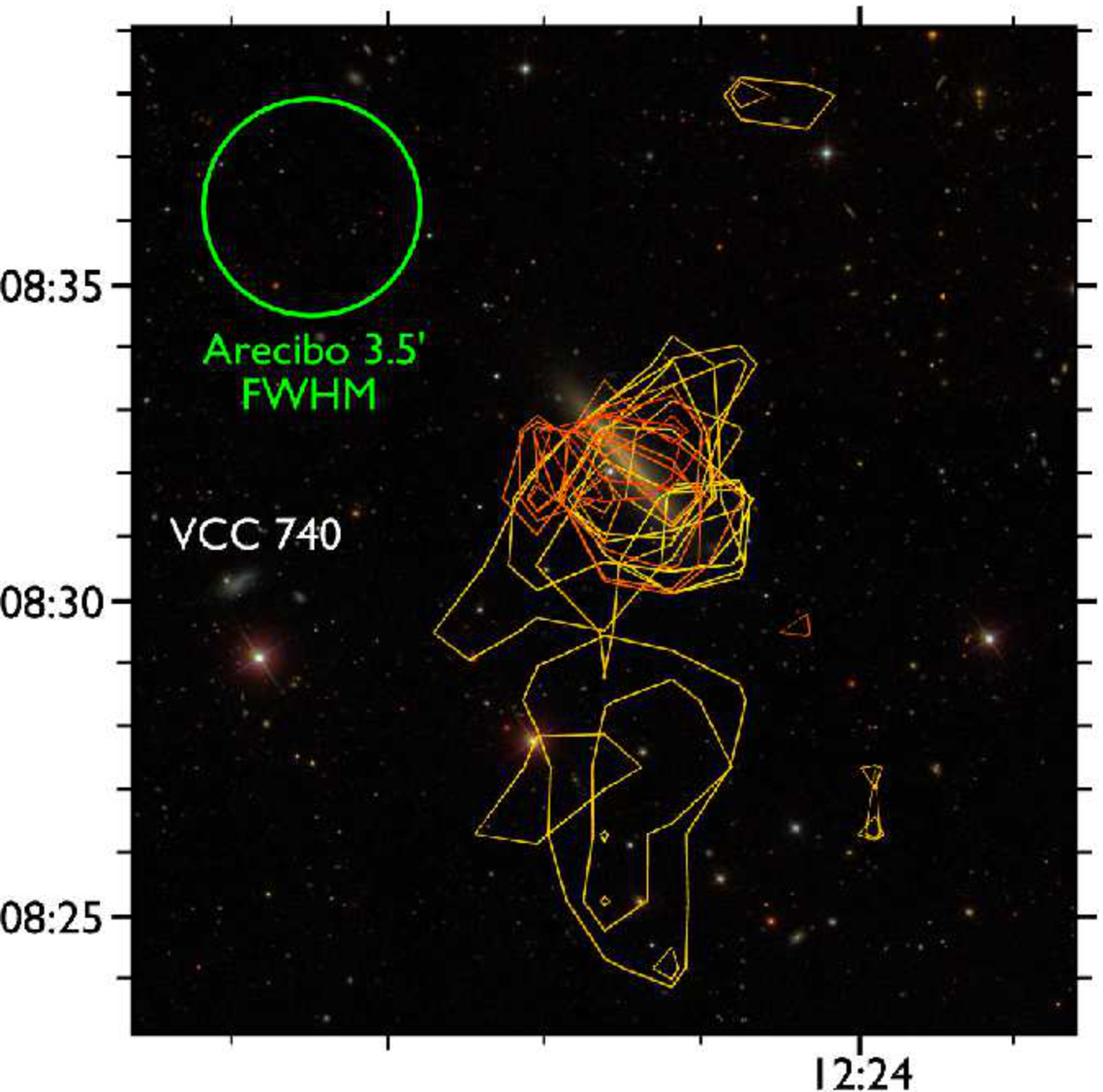}
\caption{VCC 713}
\label{fig:VCC713}
\end{figure}

VCC 713 is a bright, red spiral which is just detected by AGES and strongly \HI{} deficient. Its stream extends due south (figure \ref{fig:VCC713}), and although it is significantly extended and detected at 5$\sigma$, it is only present in a couple of channels. Given the high predicted mass lost from this galaxy and the absence of a massive stream, it might be on its second orbit so that the stream has long since dispersed.

This galaxy is shown in figure \ref{fig:VCC713} with the contour at 4$\sigma$ over the velocity range 1,012 - 1,261\,\kms{}.

\subsubsection{VCC 1725}
At 4$\sigma$ and only about 1 beam length across, this stream is a tenatative detection at best. It appears in 3-4 velocity channels, though rather more at 3$\sigma$. No other galaxies are visible nearby.

\begin{figure}
\centering
\includegraphics[width=85mm]{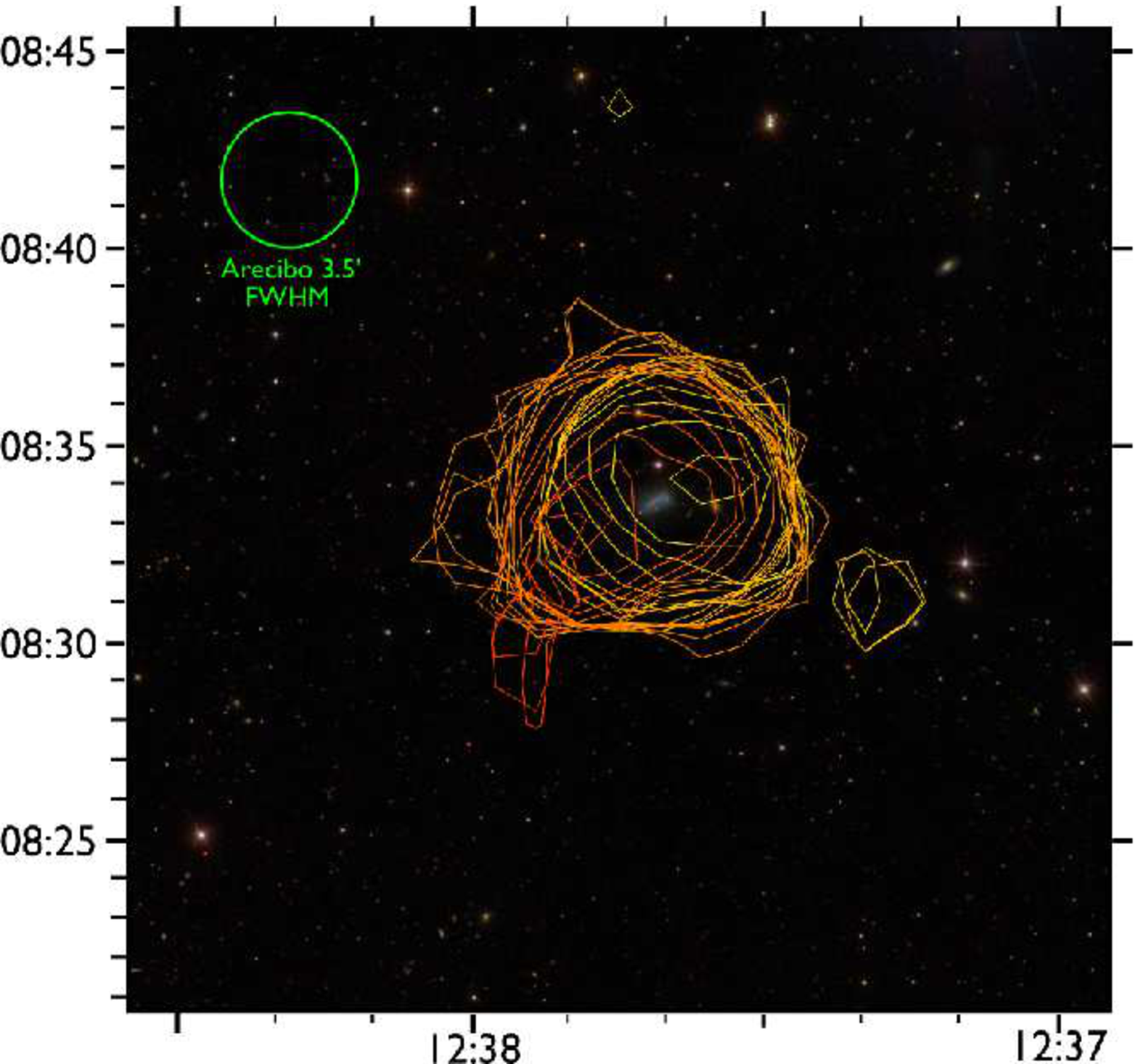}
\caption{VCC 1725}
\label{fig:VCC1725}
\end{figure}

This galaxy is shown in figure \ref{fig:VCC1725} with the contour at 4$\sigma$ over the velocity range 1,004 - 1,196\,\kms{}.

\subsubsection{GLADOS 001}
\begin{figure}
\centering
\includegraphics[width=85mm]{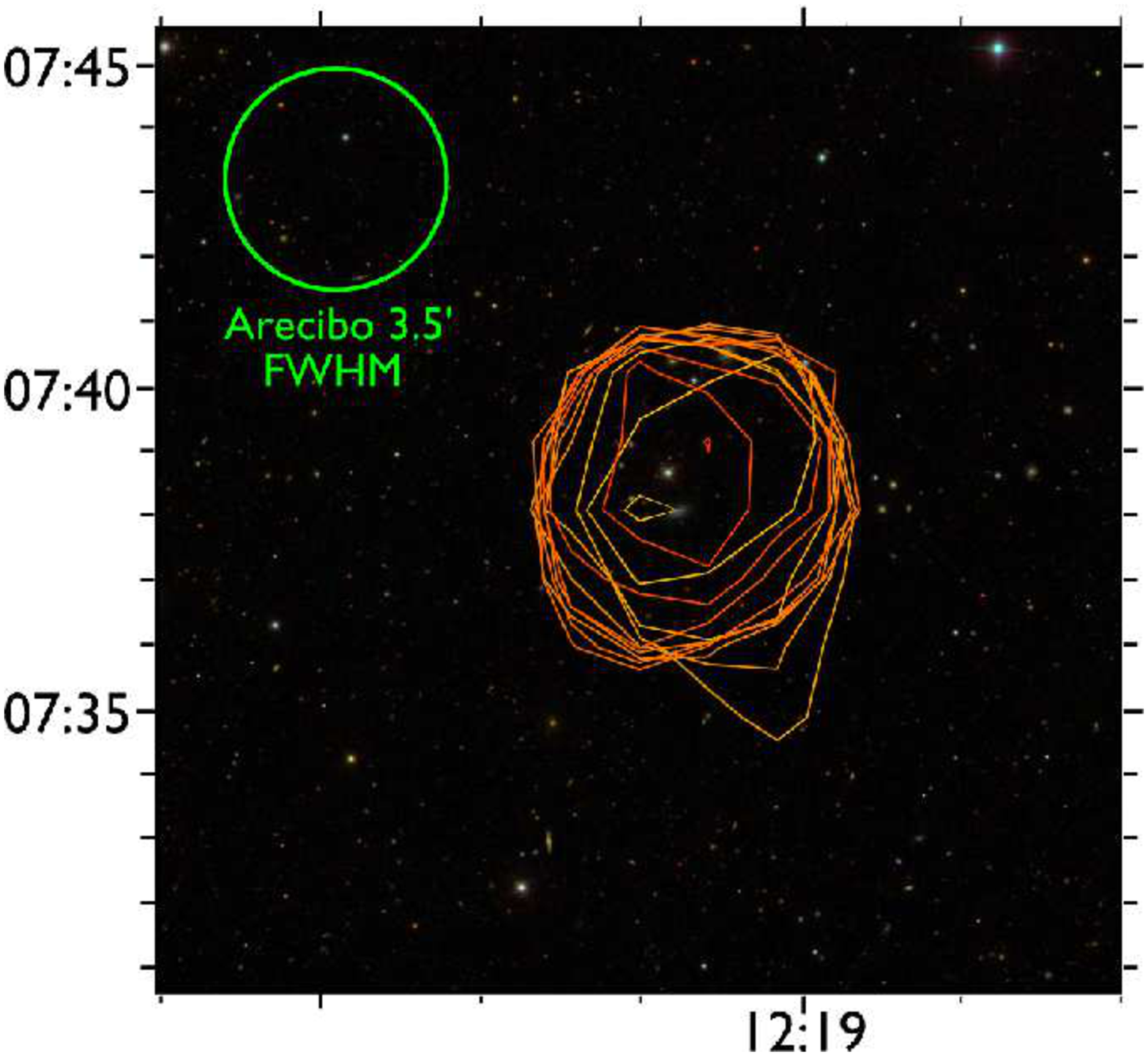}
\caption{GLADOS 001}
\label{fig:GLADOS001}
\end{figure}

The detection of the extension from this dwarf irregular is marginal at best, visible at 4$\sigma$ in only 2-3 channels (figure \ref{fig:GLADOS001}). There are no obvious galaxies that could be tidally interacting.

This galaxy is shown in figure \ref{fig:GLADOS001} with the contour at 4$\sigma$ over the velocity range 2,634 - 2,718\,\kms{}.

\subsubsection{VCC 514}
\begin{figure}
\centering
\includegraphics[width=85mm]{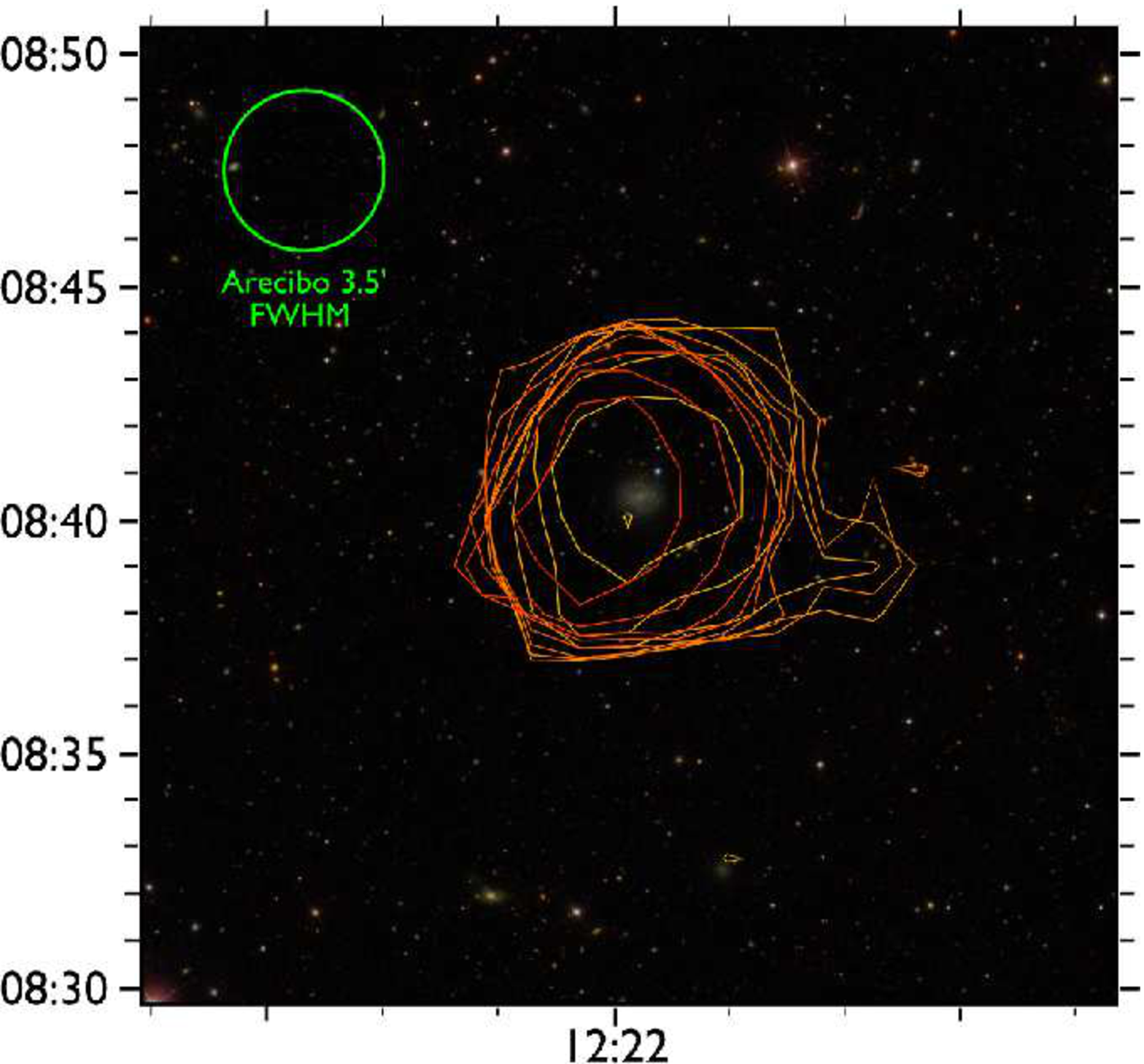}
\caption{VCC 514}
\label{fig:VCC514}
\end{figure}

This small, faint spiral has a possible extension to the west, visible at 4$\sigma$ in 4 channels (figure \ref{fig:VCC514}). No companion galaxies are visible. 

This galaxy is shown in figure \ref{fig:VCC514} with the contour at 4$\sigma$ over the velocity range 807 - 903\,\kms{}.

\subsubsection{AGESVC1 235}
AGESVC1 235 is a small blue dwarf galaxy with a possible roughly north-south extension, visible at 5$\sigma$ in 3 velocity channels. VCC 393, a disturbed spiral galaxy, is only a few arcminutes directly north but its velocity of 2,618 \kms{} is quite different to that of AGESVC1 235 (1,667 \kms{}). AGESVC1 127 is also close on the sky but it a background galaxy with a velocity $>$\,7,000 \kms{}.

\begin{figure}
\centering
\includegraphics[width=85mm]{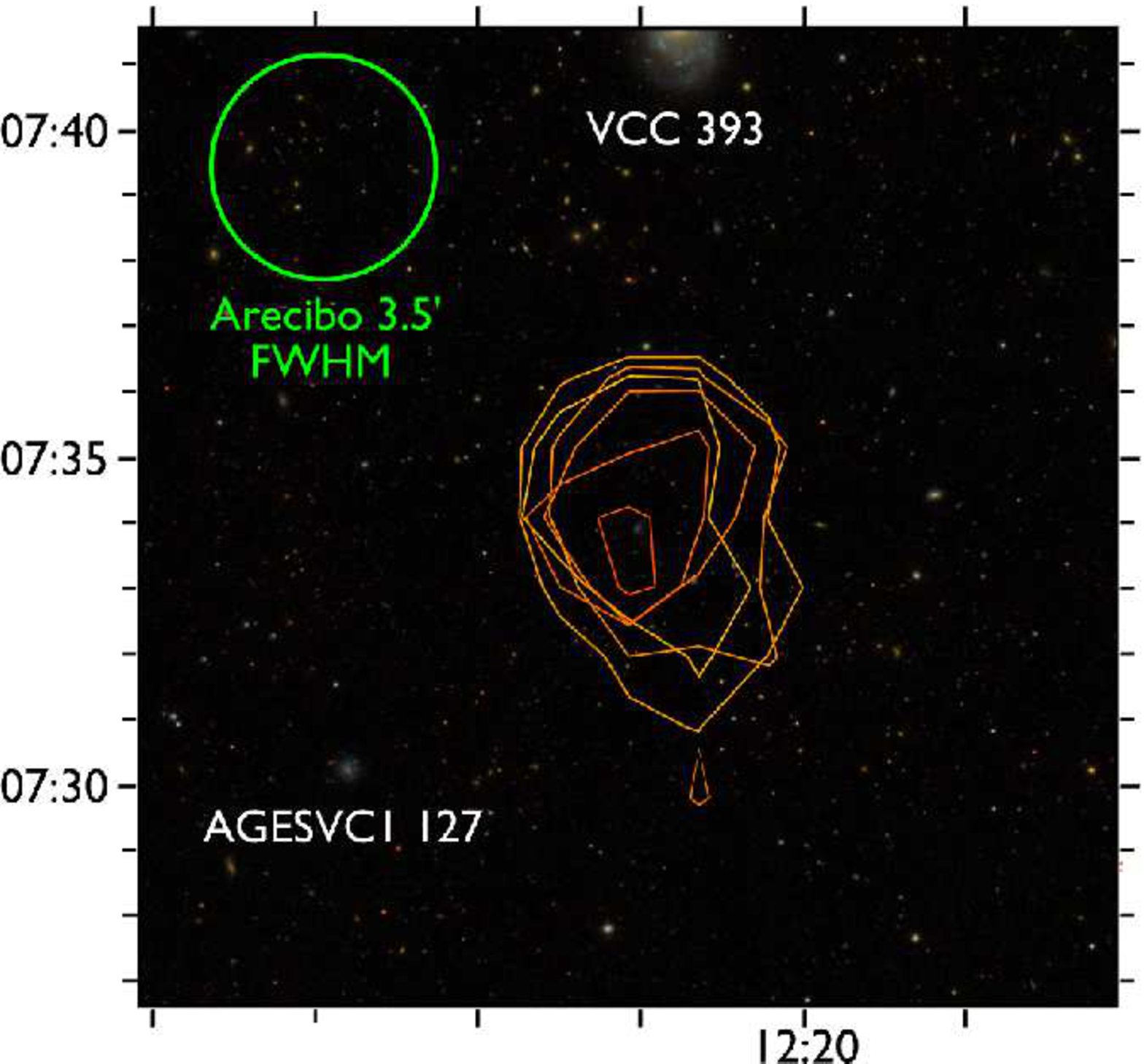}
\caption{AGESVC1 235}
\label{fig:AGESVC1235}
\end{figure}

This galaxy is shown in figure \ref{fig:AGESVC1235} with the contour at 4$\sigma$ over the velocity range 1,638 - 1,706\,\kms{}.

\subsubsection{AGESVC2 063}
\begin{figure}
\centering
\includegraphics[width=85mm]{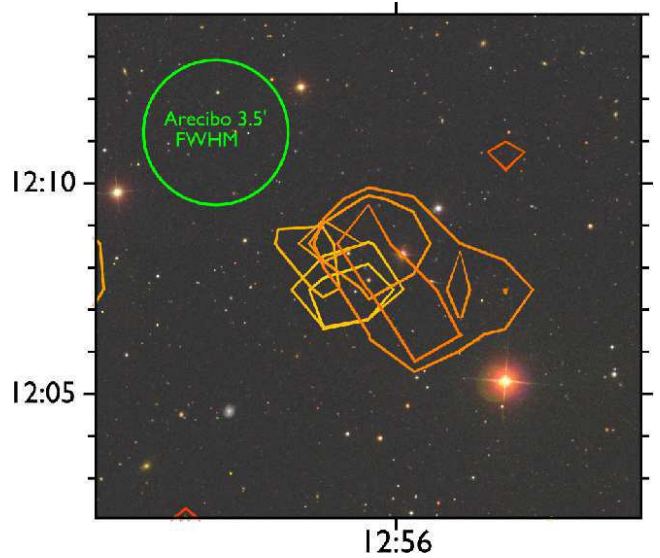}
\caption{Possible stream of the AGESVC2 063 detection. The optical counterpart is an extremely faint dwarf irregular, just visible in the centre of the image.}
\label{fig:vc2063}
\end{figure}

While the \HI{} signal itself is secure, this stream (figure \ref{fig:vc2063}) is only visible at $\sigma$ and only present over a few channels. If confirmed, this could be evidence of a dwarf galaxy experiencing ram pressure stripping, but we caution that the appearance of a stream could simply be due to noise.

This galaxy is shown in figure \ref{fig:vc2063} with the contour at 3$\sigma$ over the velocity range 1,837 - 1,926\,\kms{}.

\subsubsection{VCC 1972}
The spiral galaxy VCC 1972 is detected as AGESVC2 022. The \HI{} appears to be entirely associated with the spiral rather than the nearby elliptical galaxy VCC 1978, which is at 300\,\kms{} lower velocity. While the main extension(s) appear to point away VCC 1978, a possible weak extension heads in the opposite direction. The nature of the extended emission is somewhat ambiguous as it is unclear whether this represents a stream or noisy contours.

\begin{figure}
\centering
\includegraphics[width=85mm]{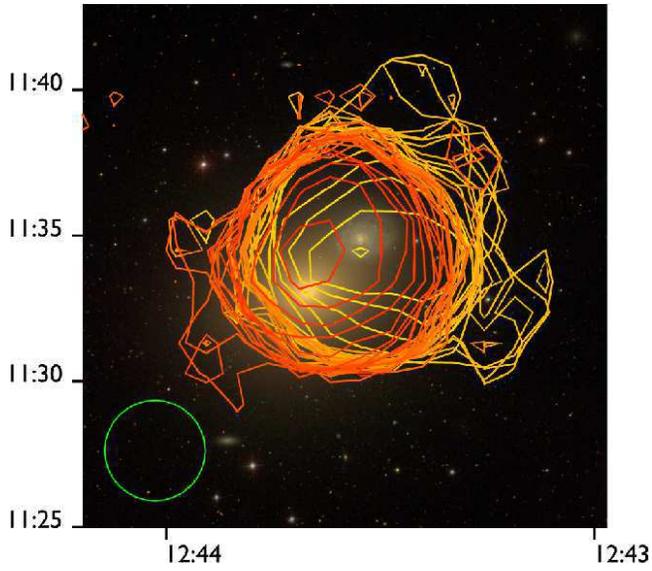}
\caption{Possible stream(s) associated with AGESVC2 022. The \HI{} detection is associated with the spiral galaxy VCC 1972, which is interacting with the giant elliptical galaxy VCC 1978.}
\label{fig:vcc1972}
\end{figure}

This galaxy is shown in figure \ref{fig:vcc1972} with the contour at 4.5$\sigma$ over the velocity range 1,205 - 1,572\,\kms{}.

\begin{thebibliography}{}

%\bibitem[\protect\citeauthoryear{Adams et al.}{2013}]{adams}
%Adams E., Giovanelli R., Haynes M. P., 2013, ApJ 768, 77
\bibitem[\protect\citeauthoryear{Auld et al.}{2006}]{auld}
Auld R., Minchin R. F., Davies J. I., et al. 2006, MNRAS, 371, 1617
%\bibitem[\protect\citeauthoryear{Barnes et al.}{2001}]{barnes}
%Barnes D. G., Staveley-Smith L., de Blok W. J. G., et al. 2001, MNRAS, 322, 486
%\bibitem[\protect\citeauthoryear{Beasley et al.}{2016}]{bee}
%Beasley M. A., Romanowsky A. J., Pota V., Mart\'{i}n Navarro I., Martinez Delgado D.; Neyer F., Deich A. L., 2016, ApJ, 819, 20
%\bibitem[\protect\citeauthoryear{Arrigoni Battaia et al.}{2012}]{batt}
%Arrigoni Battaia F., Gavazzi G., Fumagalli M., et al., 2012, A\&A, 534, 112
\bibitem[\protect\citeauthoryear{Bekki et al.}{2005}]{bekki}
Bekki K., Koribalski B. S., Kilborn V. A., 2005, MNRAS, 363, 21
\bibitem[\protect\citeauthoryear{Boselli \& Gavazzi}{2006}]{bg06}
Boselli A., Gavazzi G., 2006, PASP, 118, 517
%\bibitem[\protect\citeauthoryear{Boselli et al.}{2009}]{boss09}
%Boselli A., Gavazzi G., 2009, A\&A, 508, 201
\bibitem[\protect\citeauthoryear{Boselli et al.}{2016}]{boss16}
Boselli A., Cuillandre J. C., Fossati M., Boissier S., Bomans D., Consolandi G., Anselmi G., Cortese L., 2016, A\&A, 587, 68
%\bibitem[\protect\citeauthoryear{Boselli et al.}{2018A}]{boss18}
%Boselli A., Fossati M., Ferrarese L., et al., 2018A, A\&A, 614, 56
\bibitem[\protect\citeauthoryear{Boselli et al.}{2018B}]{boss18b}
Boselli A., Fossati M., Consolandi G., et al., 2018B, A\&A, 620, 164
\bibitem[\protect\citeauthoryear{Boselli et al.}{2018B}]{boss19}
Boselli A., Epinat B., Contini T., et al., 2019, A\&A, in press
%\bibitem[\protect\citeauthoryear{Binggeli, Sandage \& Tammann}{1985}]{bing}
%Binggeli B., Sandage A., Tammann G. A., 1985, AJ, 90, 168
%\bibitem[\protect\citeauthoryear{Broeils \& Rhee}{1997}]{brhee}
%Broeils A. H., Rhee M. H., 1997, A\&A, 324, 877
%\bibitem[\protect\citeauthoryear{Brooks et al.}{2017}]{brooks}
%Brooks A M., Papastergis E., Christensen C. R., et al., 2017, ApJ, submitted
%\bibitem[\protect\citeauthoryear{Burkhart \& Loeb}{2016}]{BL}
%Burkhart B., Loeb A., 2016, ApJ, in press
%\bibitem[\protect\citeauthoryear{Cannon et al.}{2015}]{cannon}
%Cannon, J.; Martinkus C.; Leisman L.; Haynes M.; Adams E.; Giovanelli R.; Hallenbeck G.; 2015, AJ, 149, 72
%\bibitem[\protect\citeauthoryear{Cen, Roxana Pop \& Bachall}{2014}]{cen}
%Cen R., Roxana Pop A., Bachall N. A., 2014, PNAS, 111, 7914
%\bibitem[\protect\citeauthoryear{Chengalur et al.}{1995}]{chen}
%Chengalur J. N., Giovanelli R., Haynes M. P., 1995, AJ, 109, 6
\bibitem[\protect\citeauthoryear{Chung et al.}{2007}]{chung}
Chung A., van Gorkom J. H., Kenney J. D. P., Vollmer B., 2007, ApJ, 659, 115
\bibitem[\protect\citeauthoryear{Chung et al.}{2009}]{chatlas}
Chung A., van Gorkom J. H., Kenney J. D. P., Crowl H., Vollmer B., 2009, AJ, 138, 1741
%\bibitem[\protect\citeauthoryear{Clark et al.}{2005}]{clark}
%Clark P. C., Bonnell I. A., Zinnecker H., Bate M. R., 2005, MNRAS, 359, 809
%\bibitem[\protect\citeauthoryear{Dalgarno \& McCray}{1972}]{dgmc}
%Dalgarno A., McCray R. A., 1972, ARA\&A, 10, 375
%\bibitem[\protect\citeauthoryear{Davies et al.}{2004}]{d04}
%Davies J., Minchin R., Sabatini S., van Driel W., Baes M., Boyce P., de Blok W. J. G., Disney M., 2004, MNRAS, 349, 922
%\bibitem[\protect\citeauthoryear{Davies, Davies \& Keenan}{2015}]{d15}
%Davies J. I., Davies L. J. M., Keenan O. C., 2015, MNRAS, 456, 1607
%\bibitem[\protect\citeauthoryear{Doyle et al.}{2005}]{doyle}
%Doyle M. T., Drinkwater M. J., Rohde D. J., Pimbblet K. A., Read M., Meyer M. J., Zwaan M. A.; Ryan-Weber E., et al., 2005, MNRAS, 361, 34
\bibitem[\protect\citeauthoryear{Duc et al.}{2007}]{duc07}
Duc P. A., Braine J., Lisenfeld U., Brinks E., Boquien M., 2007, A\&A, 475, 187
%\bibitem[\protect\citeauthoryear{Duc \& Bournaud}{2008}]{duc}
%Duc P. A., Bournaud F., 2008, ApJ, 673, 787
\bibitem[\protect\citeauthoryear{Duc \& Bournaud}{2008}]{duc08}
Duc P. A., Bournaud F., 2008, ApJ, 673, 787
\bibitem[\protect\citeauthoryear{Espada et al.}{2011}]{espa}
Espada D., Verdes-Montenegro L., Huchtmeier W. K., Sulentic J., Verley S., Leon S., Sabater J., 2011, A\&A, 532, 117 
\bibitem[\protect\citeauthoryear{Ferrarese et al.}{2012}]{ngvs}
Ferrarese L., C\^{o}te P., Cuillandre J. C., Gwyn S. D. J., Peng E. W., MacArthur L. A., Duc P. A., Boselli A., et al., 2012, ApJ, 200, 4
%\bibitem[\protect\citeauthoryear{Fryxell et al.}{2000}]{fryxell}
%Fryxell B., Olson K., Ricker P., Timmes F. X., Zingale M., Lamb D. Q., MacNeice P., Rosner R., et al., 2000, ApJ, 131, 273
%\bibitem[\protect\citeauthoryear{Fossati et al.}{2018}]{foss}
%Fossati M., Mendel J. T., Bosseli A., et al., 2018, A\&A, 614, 57
%\bibitem[\protect\citeauthoryear{Gavazzi et al.}{1999}]{gav3D}
%Gavazzi G., Boselli A., Scodeggio M., Pierini D., Belsole E., 1999, MNRAS, 304, 595
%\bibitem[\protect\citeauthoryear{Gavazzi et al.}{2003}]{gm}
%Gavazzi G., Boselli A., Donati A., Franzetti P., Scodeggio M., 2003, A\&A, 400, 451
%\bibitem[\protect\citeauthoryear{Gavazzi et al.}{2005}]{gv05}
%Gavazzi G., Boselli A., van Driel W., O'Neil K., 2005, A\&A, 429, 439
\bibitem[\protect\citeauthoryear{Gavazzi et al.}{2008}]{gv08}
Gavazzi G., Giovanelli R., Haynes M. P., et al., 2008, A\&A, 482, 43
%\bibitem[\protect\citeauthoryear{Gavazzi et al.}{2013}]{gv13}
%Gavazzi G., Fumagalli M., Fossati M., Galardo V., Grossetti F., Boselli A., Giovanelli R., Haynes M. P., 2013, A\&A, 553, 89
\bibitem[\protect\citeauthoryear{Gavazzi et al.}{2018}]{gv18}
Gavazzi G., Consolandi G., Gutierrez M. L., Boselli A., Yoshida M., 2018, A\&A, 618, 130
\bibitem[\protect\citeauthoryear{Giovanelli \& Haynes}{1989}]{ghcloud}
Giovanelli R., Haynes M.P., 1989, ApJ, 346, 5
\bibitem[\protect\citeauthoryear{Giovanelli et al.}{2005}]{alfalfa}
Giovanelli R., Haynes M. P., Kent B. R., Perillat P., Saintonge A., Brosch N., Catinella B., Hoffman G. L., et al., 2005, AJ, 130, 2598
%\bibitem[\protect\citeauthoryear{Gooch}{1996}]{kvis}
%Gooch R., 1996. ASP Conf., 101, 80
\bibitem[\protect\citeauthoryear{Grossi}{2008}]{gm33}
Grossi M., Giovanardi C., Corbelli E., Giovanelli R., Haynes M. P., Martin A. M., Saintonge A., Dowell J. D., 2008, A\&A, 487, 161
\bibitem[\protect\citeauthoryear{Gunn \& Gott}{1972}]{guns}
Gunn J. E., Gott R. J., 1972, ApJ, 176, 1
%\bibitem[\protect\citeauthoryear{Hallenbeck et al. }{2012}]{hallenbeck}
%Hallenbeck G., Papastergis E., Huang S., Haynes M. P., Giovanelli R., Boselli A.; Boissier S., Heinis S, et al., 2012, AJ, 144, 87
\bibitem[\protect\citeauthoryear{Haynes \& Giovanelli}{1984}]{haynes84}
Haynes M. P., Giovanelli R., 1984, AJ, 89, 7
\bibitem[\protect\citeauthoryear{Haynes \& Giovanelli}{1986}]{haynes86}
Haynes M. P., Giovanelli R., 1986, ApJ, 306, 466
%\bibitem[\protect\citeauthoryear{Haynes et al.}{2018}]{aa100}
%Haynes M. P., Giovanelli R., Kent, B. R., et al., 2018, ApJ, 861, 49
\bibitem[\protect\citeauthoryear{J\'{a}chym et al.}{2007}]{pavel1}
J\'{a}chym P., Palou\v{s} J., K\"{o}ppen J., Combes F., 2007, A\&A, 472, 5
\bibitem[\protect\citeauthoryear{J\'{a}chym et al.}{2009}]{pavel2}
J\'{a}chym P., Palou\v{s} J., K\"{o}ppen J., Combes F., 2009, A\&A, 500, 692
%\bibitem[\protect\citeauthoryear{J\'{a}chym et al.}{2014}]{pavel4}
%J\'{a}chym P., Combes F., Cortese L., Sun M., Kenney J. D., ApJ, 792, 11
\bibitem[\protect\citeauthoryear{Jaff\'{e} et al.}{2015}]{jaff}
Jaff\'{e} Y., Smith R., Candlish G. M., Poggianti B. M., Sheen Y., Verheijen M., 2015, MNRAS, 448, 1715
%\bibitem[\protect\citeauthoryear{Jones et al.}{2017}]{aa70}
%Jones, M. G., Papastergis E., Haynes M. P., Giovanelli R., 2016, MNRAS, 457, 4393
\bibitem[\protect\citeauthoryear{Keenan et al.}{2016}]{olivia}
Keenan O. C., Davies J. I., Taylor R., Minchin R. F., 2016, MNRAS, 456, 951
\bibitem[\protect\citeauthoryear{Kent et al.}{2009}]{kent09}
Kent B., Spekkens K., Giovanelli R., Haynes M., Momjian E., Cort\'{e}s J.; Hardy E., West A., 2009, ApJ, 691, 1595
\bibitem[\protect\citeauthoryear{Kent}{2010}]{kent10}
Kent B., 2010, ApJ, 725, 2333
%\bibitem[\protect\citeauthoryear{Klypin et al.}{2015}]{klypin}
%Klypin A., Karachentsev I., Makarov D., Nasonova O., 2015, MNRAS, 454, 1798
%\bibitem[\protect\citeauthoryear{Koda et al.}{2015}]{koda}
%Koda J., Yagi M., Yamanoi H., Komiyama Y., 2015, ApJ, 807, 2
\bibitem[\protect\citeauthoryear{K\"{o}ppen et al.}{2018}]{koppen}
K\"{o}ppen J., J\'{a}chym P., Taylor R., Palou\v{s} J., 2018, MNRAS, 479, 4367
\bibitem[\protect\citeauthoryear{Koopmann et al.}{2008}]{k08}
Koopmann R. A., Giovanelli R., Haynes M. P., et al. 2008, ApJ, 682, L85
%\bibitem[\protect\citeauthoryear{Koyama \& Inutsuka}{2002}]{koyama}
%Koyama H., Inutsuka S., 2002, ApJ, 564, 97
%\bibitem[\protect\citeauthoryear{Koyama \& Inutsuka}{2000}]{koyama00}
%Koyama H., Inutsuka S., 2000, ApJ, 532, 980
%\bibitem[\protect\citeauthoryear{Leisman et al.}{2017}]{aaudgs}
%Leisman L., Haynes M. P., Janowiecki S., 2017, ApJ, submitted
%\bibitem[\protect\citeauthoryear{Leroy et al.}{2008}]{leroy}
%Leroy A. K., Walter F., Brinks E., Bigiel F., de Blok W. J. G., Madore B., Thornley M. D., 2008, AJ, 136, 2782
%\bibitem[\protect\citeauthoryear{Larson, Tinsely \& Caldwell}{1980}]{starve}
%Larson R. B., Tinsely B. M., Caldwell C. N., 1980, ApJ, 237, 692 
%\bibitem[\protect\citeauthoryear{Mei et al.}{2007}]{Mei}
%Mei S., Blakeslee J. P., C\^{o}te P., Tonry J. L., West M. J., Ferrarese L., Andr\'{e}as J., Peng E. W., Andr\'{e} A., Merritt D., 2007, ApJ, 655, 144
%\bibitem[\protect\citeauthoryear{Mihos et al.}{2015}]{mihos}
%Mihos J., Durrell P. R., Ferrarese L., Feldmeier J. J., C\^{o}te P.; Peng E. W., Harding P., Liu C., et al., 2015, ApJ, 809, 21
\bibitem[\protect\citeauthoryear{Minchin et al.}{2007}]{m07}
Minchin R., Davies J. I., Disney M., Grossi M., Sabatini S., Boyce P., Garcia D., Impey C., et al., 2007, ApJ, 670, 1056
\bibitem[\protect\citeauthoryear{Minchin et al.}{2010}]{m10}
Minchin R. F., Momjian E., Auld R., Davies J. I., Valls-Gabaud D., Karachentsev I. D., Henning P. A., O'Neil K. L., et al., 2010, MNRAS, 140, 1093
\bibitem[\protect\citeauthoryear{Minchin et al.}{2016}]{m16}
Minchin R. F., Auld R., Davies J. I., et al., 2016, MNRAS, 455, 3430
%\bibitem[\protect\citeauthoryear{Moore et al.}{1995}]{mooreharas}
%Moore B., Katz N., Lake G., Dressler A., Oemler A., 1996, Nature, 379, 613
%\bibitem[\protect\citeauthoryear{Moore et al.}{1999}]{moore}
%Moore B., Ghigna S., Governato F., Lake G., Quinn T., Stadel J., Tozzi P., 1999, ApJ, 524, 19
%\bibitem[\protect\citeauthoryear{Mu\~{n}oz et al.}{2015}]{munoz}
%Mu\~{n}oz R. P., Eigenthaler P., Puzia T. H, Taylor M. A.; Ordenes-Brice\~{n}o Y.; Alamo-Mart\'{i}nez K., Ribbeck K. X.; \'{A}ngel S., 2015, ApJ, 813, 15
\bibitem[\protect\citeauthoryear{Minchin et al.}{2019}]{m19}
%Minchin R. F., Taylor R., K\"{o}ppen J., Davies J. I., van Driel W., Keenan 0., 2019, 158, 121
\bibitem[\protect\citeauthoryear{Noordermeer et al.}{2005}]{noord}
Noordermeer E., van der Hulst J. M., Sancisi R., Swaters R. A., van Albada T. S., 2005, A\&A, 442, 137
\bibitem[\protect\citeauthoryear{Oosterloo et al.}{2005}]{oo05}
Oosterloo T., van Gorkom J., 2005, A\&A, 437, 19
\bibitem[\protect\citeauthoryear{Pisano, Wilcots \& Liu}{2002}]{pwl}
Pisano D. J., Wilcots E. M., Liu C. T., 2002, ApJ, 142, 161
\bibitem[\protect\citeauthoryear{Portas et al.}{2011}]{port}
Portas A., Scott T. C., Brinks E., et al., 2011, ApJ, 739, 27
\bibitem[\protect\citeauthoryear{Putman et al.}{2002}]{put}
Putman M. E., de Heij V., Staveley-Smith L., Braun R., Freeman K. C., Gibson B. K., Burton W. B., Barnes D. G., et al., 2002, AJ, 123, 873
%\bibitem[\protect\citeauthoryear{Roberts et al.}{2004}]{roberts}
%Roberts S., Davies J., Sabatini S., van Driel W., O'Neil K., Baes M., Linder S., et al., 2004, MNRAS, 352, 478
\bibitem[\protect\citeauthoryear{Ram\'{i}rez-Moreta et al.}{2018}]{ramir}
Ram\'{i}rez-Moreta P., Verdes-Montenegro L., Blasco-Herrera, 2018, A\&A, 619, 163
\bibitem[\protect\citeauthoryear{Rhee et al.}{2017}]{rhee}
Rhee J., Smith R., Choi H., Yi S., Jaff\'{e} Y., Candlish G., S\'{a}nchez-J\'{a}nssen R., 2017, ApJ, 843, 128 
\bibitem[\protect\citeauthoryear{Roediger \& Br\"{u}ggen}{2008}]{roe08}
Roediger E., Br\"{u}ggen M., 2008, MNRAS, 388, 465
\bibitem[\protect\citeauthoryear{Ruppen}{1999}]{rupp}
Ruppen M. P., 1999, ASPC, 180, 229
%\bibitem[\protect\citeauthoryear{Sabatini et al.}{2003}]{sabatini}
%Sabatini S., Davies J., Scaramella R., Smith R., Baes M., Linder S. M., Roberts S., Testa V., 2003, MNRAS, 341, 981
%\bibitem[\protect\citeauthoryear{Saintonge}{2007}]{saintonge}
%Saintonge A., 2007, AJ, 133, 2087
%\bibitem[\protect\citeauthoryear{Sawala et al.}{2015}]{sawala}
%Sawala T., Frenk C. S., Fattahi A., et al., 2016, MNRAS, 457, 1931
%\bibitem[\protect\citeauthoryear{Schindler, Binggeli \& B\"{o}hhringer}{1999}]{schindler}
%Schindler S., Binggeli B., B\"{o}hhringer H., 1999, A\&A, 343, 420
%\bibitem[\protect\citeauthoryear{Schure et al.}{2009}]{schure}
%Schure K. M., Kosenko D., Kaastra J. S., Keppens R., Vink J., 2009, A\&A, 508, 751
%\bibitem[\protect\citeauthoryear{Shibata et al.}{2001}]{shib}
%Shibata R., Matsushita K., Yamasaki N. Y., Ohashi T., Ishida T., Kikuchi K., B\"{o}hringer H., Matsumoto H., 2001, ApJ, 548, 228 
\bibitem[\protect\citeauthoryear{Scott et al.}{2012}]{scott}
Scott T. C., Sengupta C., Verdes Montenegro L., et al., 2014, A\&A, 567, 56
\bibitem[\protect\citeauthoryear{Scott et al.}{2018}]{greatscott}
Scott T. C., Brinks E., Cortese L., Boselli A., Bravo-Alfaro H., 2018, MNRAS, in press
\bibitem[\protect\citeauthoryear{Solanes, Giovanelli \& Haynes}{1996}]{solanes}
Solanes J. M., Giovanelli R., Haynes M. P., 1996, ApJ, 461, 609
\bibitem[\protect\citeauthoryear{Solanes et al.}{2001}]{solanes01}
Solanes J. M., Manrique A., Garc\'{i}a-G\'{o}mez C., Gonz\^{a}lez-Casado G.,  Giovanelli R., Haynes M. P., 2001, ApJ, 548, 97
\bibitem[\protect\citeauthoryear{Sorgho et al.}{2017}]{sorgho}
Sorgho A., Hess K., Carignan C., Oosterloo T., 2017, MNRAS, 464, 530
%\bibitem[\protect\citeauthoryear{Sutherland \& Dopita}{1993}]{cooling}
%Sutherland R. S., Dopita M. A., 1993, ApJ, 88, 253
%\bibitem[\protect\citeauthoryear{Sterbnerg, McKee \& Wolfire}{2002}]{stern}
%Sternberg A., McKee C. F., Wolfire M. G., 2002, ApJ, 143, 419
%\bibitem[\protect\citeauthoryear{Sun et al.}{2010}]{sun}
%Sun M., Donahue M., Roediger E., Nulsen P. E. J., Vorr G. M., Sarazin C., Forman W., Jones C., 2010, ApJ, 708, 946
\bibitem[\protect\citeauthoryear{Swaters et al.}{2002}]{swat}
Swaters R. A., van Albada T. S., van der Hulst J. M., Sancisi R., 2002, A\&A, 390, 829
\bibitem[\protect\citeauthoryear{Taylor}{2006}]{stilts}
Taylor, M. B., 2006, ASPC, 351, 666
%\bibitem[\protect\citeauthoryear{Taylor}{2010}]{mythesis}
%Taylor, Rh., PhD thesis, 2010, Department of Physics and Astronomy, Cardiff University, 5 The Parade, Cardiff, Glamorgan, CF24 3YB, Wales, U.K.
\bibitem[\protect\citeauthoryear{Taylor et al.}{2012}]{me12}
Taylor R., Auld R., Davies J. I., Minchin R. F., 2012, MNRAS, 423, 787
\bibitem[\protect\citeauthoryear{Taylor et al.}{2013}]{me13}
Taylor R., Minchin R. F., Herbst H., Davies J. I., Rodriguez R., Vasquez C., 2013, MNRAS, 428, 459
\bibitem[\protect\citeauthoryear{Taylor et al.}{2014}]{me14}
Taylor R., Minchin, R. F., Herbst H., Smith R., 2014, MNRAS, 442, 46
\bibitem[\protect\citeauthoryear{Taylor}{2015}]{me15}
Taylor R., 2015, A\&C, 13, 67
\bibitem[\protect\citeauthoryear{Taylor et al.}{2016}]{me16}
Taylor R., Davies J. I., J\'{a}chym P., Keenan O., Minchin R. F., Palou\v{s} J., Smith R., W\"{u}nsch R., 2016, MNRAS, 461, 300
\bibitem[\protect\citeauthoryear{Taylor et al.}{2017}]{me17}
Taylor R., Davies J. I., J\'{a}chym P., Keenan O., Minchin R. F., Palou\v{s} J., Smith R., W\"{u}nsch R., 2017, MNRAS, 467, 3648
\bibitem[\protect\citeauthoryear{Taylor, W\"{u}nsch \& Palou\v{s}}{2018}]{me18}
Taylor R., W\"{u}nsch R., Palou\v{s} J., 2018, MNRAS, in press
\bibitem[\protect\citeauthoryear{Tonnesen \& Bryan}{2010}]{ton10}
Tonnesen S., Bryan G. L., 2010, ApJ, 709, 1203
%\bibitem[\protect\citeauthoryear{van der Burg}{2016}]{van16}
%van der Burg R. F. J., 2016, Muzzin A., Hoekstra H., 2016, A\&A, 590, 20
%\bibitem[\protect\citeauthoryear{van Dokkum et al.}{2015}]{vandok}
%van Dokkum P. G., Abraham R., Merritt A., Zhang J., Geha M., Conroy C., 2015, ApJ, 798, 45
%\bibitem[\protect\citeauthoryear{van Dokkum et al.}{2016}]{d44}
%van Dokkum P., Abraham R., Brodie J., et al., 2016, ApJ, 828, 6
%\bibitem[\protect\citeauthoryear{Villaescusa-Navarro et al.}{2016}]{vilnar}
%Villaescusa-Navarro F., Planelles S., Borgani S., Viel M., Rasia E., Murante G., Dolag K., Steinborn L K., et al., 2016, MNRAS, 456, 3553
%\bibitem[\protect\citeauthoryear{Tully et al.}{1996}]{virgocross}
%Tully R. B., Verheijen M. A. W., Pierce M. J., Huang J., Wainscoat R. J., 1996, AJ, 112, 2471 
\bibitem[\protect\citeauthoryear{Toomre \& Toomre}{1972}]{tooms}
Toomre A., Toomre J., 1972, ApJ, 178,623
%\bibitem[\protect\citeauthoryear{van Driel \& van Woerden}{1989}]{vans}
%van Driel W., van Woerden H., 1989, A\&A, 225, 317
\bibitem[\protect\citeauthoryear{Verdugo et al.}{2015}]{verd}
Verdugo C., Combes F., Dasyra K., Salom\'{e} P., Braine J., 2015, A\&A, 582, 6
\bibitem[\protect\citeauthoryear{Vollmer et al.}{2001}]{vol01}
Vollmer B., Cayatte V., Balkowski C., Duschl W. J., 2001, ApJ, 561, 708
\bibitem[\protect\citeauthoryear{Vollmer et al.}{2007}]{vol07}
Vollmer B., Hutchmeier W., 2007, A\&A, 462, 93
%\bibitem[\protect\citeauthoryear{Whiting}{2012}]{duchamp}
%Whiting M. T., 2012, MNRAS, 421, 3242
%\bibitem[\protect\citeauthoryear{Wolfire et al.}{1995}]{wolf}
%Wolfire M. G., Hollenbach D., McKee C. F., Tielens A. G. G. M., Bakes E. L. O., 1995, ApJ, 443, 152
\bibitem[\protect\citeauthoryear{Wang et al.}{2016}]{wang}
Wang J., Koribalski B. S., Serra P., et al., 2016, MNRAS, 460, 2143
%\bibitem[\protect\citeauthoryear{Yagi et al.}{2010}]{yagi10}
%Yagi M., Yoshida M., Komiyama Y., 2010, AJ, 140, 1814
%\bibitem[\protect\citeauthoryear{Yagi et al.}{2017}]{yagi}
%Yagi M., Yoshida M., Gavazzi G., Komiyama Y., Kashikawa N., Okamura S., 2017, ApJ, 839, 65
%\bibitem[\protect\citeauthoryear{Zwaan et al.}{1995}]{zwaan}
%Zwaan M. A., van der Hulst J. M., de Blok W. J. G., McGaugh S. S., 1995, MNRAS, 273, 35
\bibitem[\protect\citeauthoryear{Yun et al.}{2019}]{yun}
Yun K., Pillepich A., Zinger E., et al., 2019, MNRAS, 483, 1082
%\bibitem[\protect\citeauthoryear{Zabel et al.}{2018}]{zab}
%Zabel N., Davis T. A., Smith M. W. L., et al., 2018, MNRAS, in press

\end{thebibliography}
\end{document}